\documentclass[showpacs, oneside, twocolumn, prl, amsmath, amssymb, nofootinbib, superscriptaddress]{revtex4-1}

\usepackage{cases}
\usepackage{amsmath}
\usepackage{amssymb}
\usepackage{amsfonts}
\usepackage{amssymb}
\usepackage{dcolumn}
\usepackage{bm}
\usepackage{bbm}
\usepackage{graphicx}
\usepackage{xcolor}
\usepackage{array}
\usepackage{subfigure}
\usepackage{hyperref}
\usepackage{wasysym}

\newcommand{\be}{\begin{equation}}
\newcommand{\ee}{\end{equation}}
\newcommand{\ba}{\begin{eqnarray}}
\newcommand{\ea}{\end{eqnarray}}

\newcommand{\gsim}{\mathrel{\hbox{\rlap{\lower.55ex \hbox {$\sim$}}
                   \kern-.3em \raise.4ex \hbox{$>$}}}}
\newcommand{\lsim}{\mathrel{\hbox{\rlap{\lower.55ex \hbox {$\sim$}}
                   \kern-.3em \raise.4ex \hbox{$<$}}}}

\renewcommand{\(}{\left(}
\renewcommand{\)}{\right)}

\def\be{\begin{equation}}
\def\ee{\end{equation}}
\newcommand{\bea}{\begin{eqnarray}}
\newcommand{\eea}{\end{eqnarray}}
\def\fnl{f_{\rm NL}}

\newcommand{\mpl}{M_{\rm Pl}}

\hypersetup{colorlinks=true,
            breaklinks=true,
            pdfstartview=Fit,
            linkcolor=blue,
            citecolor=blue,
            urlcolor=blue}

\begin{document}
\title{The UV Sensitivity of Axion Monodromy Inflation}

\author{Enrico Pajer}
\email{enrico.pajer@gmail.com}
\affiliation{Centre for Theoretical Cosmology, Department of Applied Mathematics and Theoretical Physics,
University of Cambridge,
Wilberforce Road, Cambridge, CB3 0WA, U.K.}

\author{Dong-Gang Wang}
\email{dgw36@cam.ac.uk}
\affiliation{Centre for Theoretical Cosmology, Department of Applied Mathematics and Theoretical Physics,
University of Cambridge,
Wilberforce Road, Cambridge, CB3 0WA, U.K.}

\author{Bowei Zhang}
\email{bz287@cam.ac.uk}
\affiliation{Centre for Theoretical Cosmology, Department of Applied Mathematics and Theoretical Physics,
University of Cambridge,
Wilberforce Road, Cambridge, CB3 0WA, U.K.}

\begin{abstract}

{We revisit axion monodromy inflation in the context of {UV-inspired models} and point out that its cosmological observables are sensitive to heavy fields with masses far above the Hubble scale, such as the moduli of flux compactifications.} By studying a string-inspired two-field extension of axion monodromy {with a small turning rate}, we reveal that the oscillatory modulation of the axion potential leads to continuous excitation of heavy fields during inflation when the modulation frequency exceeds the field masses. This finding challenges the conventional single-field description, heavy moduli cannot be simply integrated out. Using a full bootstrap analysis, we demonstrate that this mechanism produces cosmological collider signals that bypass the usual Boltzmann suppression for heavy masses. Specifically, we identify detectably large signatures of heavy moduli in the primordial bispectrum, offering a promising avenue for probing high-energy physics through cosmological observations.

\end{abstract}


\maketitle

{\it Introduction}--
How sensitive are inflationary correlators to UV physics? The answer might be discouraging if we take a look at most of the UV-complete theories of inflation. For decades, one main focus of string cosmology has been to achieve inflation with a single active degree of freedom in 4D \cite{Baumann:2014nda,Cicoli:2023opf}. In most constructions, a large number of moduli fields arise from (flux) compactification. These moduli carry information about the UV theory, for example by encoding the geometry of extra dimensions. However, after moduli stabilization, they are supposed to be decoupled from the low-energy theory. Only in certain circumstances, integrating out these heavy states can lead to reduced sound speed and sizable self-interaction of the inflaton, which generate the equilateral-type non-Gaussianity in cosmological correlators \cite{Tolley:2009fg, Achucarro:2010da, Baumann:2011su,Achucarro:2012sm}.

Another perspective towards the UV sensitivity of inflation is provided by the cosmological collider physics ~\cite{Chen:2009zp, Baumann:2011nk, Noumi:2012vr, Arkani-Hamed:2015bza}. 
In that setup, signatures of massive particles during inflation appear as squeezed-limit oscillations of the scalar bispectrum. For heavy particles with $m\gg H$, the signals are suppressed by a Boltzmann factor $e^{-\pi m/H}$, and so this channel is sensitive only to extra fields with masses of $\mathcal{O}(H)$. {An additional mechanism is needed to enhance these signals. Two possibilities are the chemical potential proposal \cite{Wang:2019gbi, Bodas:2020yho, Tong:2022cdz} and the effective field theory (EFT) with small sound speeds \cite{Lee:2016vti,Pimentel:2022fsc,Jazayeri:2022kjy,Wang:2022eop} {or both \cite{Jazayeri:2023kji}}. A third possibility, which will be realized in our model, is the presence of features in the potential \cite{Chen:2022vzh}.}

In this letter, we re-examine the UV sensitivity in one class of models arising from stringy embeddings -- the axion monodromy inflation \cite{Silverstein:2008sg,McAllister:2008hb,Flauger:2009ab,Berg:2009tg,Kaloper:2011jz}. As one of the most successful examples of string inflation, this model breaks the discrete shift symmetry of an axion by introducing a monodromy, namely the axion potential becomes multivalued. Thus in a controllable way, the 4D effective theory provides a successful realization of large field inflation with a sub-Planckian axion decay constant.
Meanwhile, the discrete symmetry of the axion allows periodic modulations of the slow-roll potential which lead to oscillations in the background evolution. Within single field inflation, the oscillatory behaviour generates characteristic signals, namely oscillatory corrections to the power spectrum and also resonant non-Gaussianity in the primordial bispectrum \cite{Chen:2008wn,Flauger:2009ab,Flauger:2010ja,Chen:2010bka,Behbahani:2011it,Leblond:2010yq,Cabass:2018roz} (also see \cite{DuasoPueyo:2023viy,Creminelli:2024cge} for recent discussions).

Like any other string model, the full description of axion monodromy contains many heavy fields, such as the moduli of the compactification. Naively, these can be stabilised and should not affect the low-energy single-clock effective theory {because they are heavy and we have a small turning rate}. However, for axion monodromy, this is more subtle because background oscillations introduce a new energy scale (see \cite{Dong:2010in,Flauger:2016idt,Pedro:2019klo,Chen:2022vzh,Bhattacharya:2022fze,Chakraborty:2019dfh} for previous studies on the effects of heavy physics). Can we still integrate out these heavy fields? 
Or in other words, what is the regime of validity for the single field EFT? Are there UV-sensitive signatures of heavy fields in cosmological correlators?

In this work we attempt to answer these questions by studying minimal but realistic {UV-inspired} models of axion monodromy. {As expected from the general analysis of \cite{Chen:2022vzh},} we find that due to the resonance between the oscillatory couplings and quantum field fluctuations, the system can become sensitive to heavy moduli when the axion oscillates at a sufficiently high frequency. 
As a consequence, the adiabaticity condition for effective single field descriptions in \cite{Cespedes:2012hu, Achucarro:2012yr} can be violated, which leads to a continuous production of heavy moduli. {All the inflationary trajectories we consider have a small turning rate}. We compute the full primordial bispectrum using the bootstrap method, and show that the resonance removes the familiar Boltzmann suppression of the cosmological collider signals for heavy masses. {This enhancement is unrelated to the turning rate and had been noticed previously in the squeezed limit in \cite{Chen:2022vzh,Pinol:2023oux,Werth:2023pfl}.}

\vskip6pt

{\it Axion monodromy revisited}--
Let's consider a concrete construction of string inflation by highlighting generic features of compactifications. 
The standard starting point is the dimensional reduction of 10D supergravity, $\mathcal{M}_{10}\rightarrow \mathcal{M}_4\times X_6$. The compact extra dimensions $X_6$ give rise to a large number of fields in the 4D effective theory, including axions and moduli. An axion $\theta$ arises from the integration of a gauge potential over nontrivial cycles in $X_6$. The continuous shift symmetry of $\theta$ is broken into a discrete one by non-perturbative effects. The kinetic term  is given by $\frac{1}{2}f_\theta^2 (\partial_\mu \theta)^2$, with $f_\theta$ the axion decay constant.
Moduli fields include the radion modulus, which controls the volume of the extra dimensions $\mathcal{V}$. For an isotropic $X_6$ with characteristic length $L$, $\mathcal{V} \propto L^6 \propto \exp{(\sqrt{3}\rho / \mpl)}$, where $\rho$ is the canonically normalized field describing volume fluctuations. After dimensional reduction, constants of the 4D theory, such as the Planck mass, are functions of the radion modulus. This is true also for the axion decay constant. A specific example is given as 
$ f_\theta^2 \propto \mathcal{V}/L^4= f^2 \exp{({\rho}/{\sqrt{3}\mpl})}$~\cite{Silverstein:2008sg,McAllister:2008hb},
where $f$ is the stabilised value at $\rho=0$.
Thus the kinetic term from the UV provides a universal coupling between axion and moduli fields. 
\footnote{In general, other types of interactions may arise in the potential, which can generate oscillating corrections to moduli masses \cite{Dong:2010in,Flauger:2016idt,Pedro:2019klo,Bhattacharya:2022fze}. We neglect their effects for simplicity and highlight the role of kinetic mixings in this work.}

\begin{figure}[t]
\centering
\includegraphics[width=0.7\linewidth]{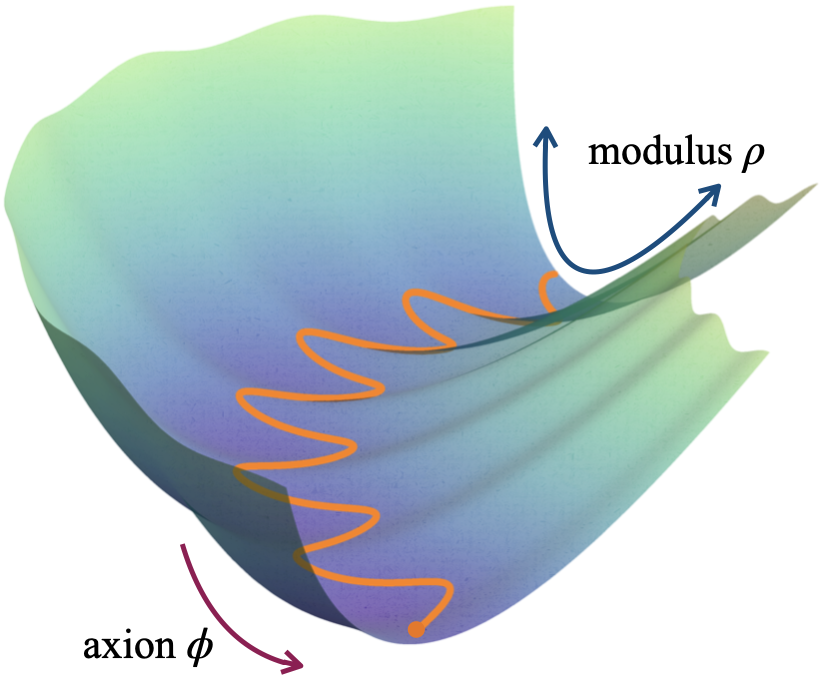}
\caption{A sketch of the axion monodromy potential with a heavy modulus field. The orange curve corresponds to the background trajectory of the inflaton with oscillations driven by the axion periodic modulation {(the wiggles have been enhanced for visibility)}.}
\label{fig:axion}
\end{figure}

To write down the 4D effective theory, we introduce the canonically normalized axion field $\phi \equiv f \theta$.  The following simple Lagrangian captures the important features of the string construction of axion monodromy
\be \label{theory}
\mathcal{L} = -\frac{1}{2} e^{\rho/\Lambda}(\partial \phi)^2 - \frac{1}{2} (\partial \rho)^2 - V(\phi, \rho) ,
\ee
where we have  introduced  $\Lambda \lesssim \mpl$ to control the coupling strength. When $\Lambda \sim \mpl$ we return to the specific example above. 
{This axion-modulus coupling can become stronger, e.g. in geometries with hierarchically different volumes \cite{Balasubramanian:2005zx}.}
The potential takes the  form
\be
 V(\phi, \rho) = V_{\rm sr}(\phi) +   A^4 \cos \left(\frac{\phi}{f}\right) + W(\rho) .
\ee
Here $V_{\rm sr}$ is a potential for the axion $\phi$ coming from monodromy, which we assume satisfies the usual slow-roll conditions. The periodic term arises from the non-perturbative instanton effects. This makes it natural for $A$ to be smaller than $V_{\rm sr}$. We also assume that $\rho$ is the lightest modulus, and that it is stabilised around $\rho=0$, with a mass $m^2 \equiv W''(\rho)\gg H^2$.  
One example of this potential is shown in Figure \ref{fig:axion}.
We emphasize that the Lagrangian \eqref{theory} is expected in any UV-completions of axion monodromy.
The common lore is that the modulus field can be seen as decoupled, and the low-energy theory reduces to single field inflation. In the following, we shall re-examine this expectation.

As the axion field develops a time-dependent background $\dot\phi_0\neq 0 $,
the frequency of axion oscillation naturally arises, $\omega \equiv \dot\phi_0/f$. For later convenience, let us specify the parametric regime of interest 
\be \label{conditions}
\alpha \equiv \frac{\dot\phi_0}{Hf} \gg 1~,~~~~ b_* \equiv \frac{A^4}{V_{\rm sr}' f} \ll 1~,~~~~ \Lambda \gg \frac{\dot\phi_0}{\sqrt{\alpha}H}~.
\ee
The first two conditions are inherited from the single field axion monodromy and correspond to a superHubble frequency of oscillations and the monotonicity of the potential. The last condition is peculiar to our two-field extension and restricts the oscillations in the modulus direction. It is needed for a controlled computation. Finally, the stability of the modulus requires $W''\gg \dot\phi_0^2/\Lambda^2$.

\vskip6pt

{\it The wiggly trajectory}--
Next, we take a look at the background dynamics of the two-field system \eqref{theory}. As
the oscillatory modulation is assumed to be small, it can be seen as a perturbation of the slow-roll evolution\footnote{We also expect oscillatory modulation of the Hubble parameter $H=H_0+H_1$, but including $H_1$ does not affect our analysis, so we simply use the constant piece. See \cite{ResonantCollider} for more details.} $\phi_B=\phi_0+\phi_1$ and $\rho_B=\rho_0+\rho_1$.  The $0$th order solution of the background equations is simply given by the slow-roll result 
$\dot\phi_0 = - {V_{\rm sr}'}/{3H}$, with $3H^2\mpl^2 = V_{\rm sr}(\phi_0)$, and we  introduce the slow-roll parameter $\epsilon_0 \equiv {\dot{\phi}_0^2}/{(2H^2M_{\mathrm{Pl}}^2)}$.
Meanwhile we also have the centrifugal force equation $\dot\phi_0^2 /(2\Lambda)=  W'(\rho_0)$ that stabilises the modulus at $\rho_0 =0$.

The first order quantities $\phi_1$ and $\rho_1$ can be obtained by imposing the conditions in \eqref{conditions}.
The trick is to notice that we are interested in high-frequency oscillations with $\alpha\gg1$, so that terms with higher time derivatives are more dominant. 
The equation for $\phi_1$ becomes the one for a driven oscillator $\ddot\phi_1 \simeq b_*  V_{\rm sr}' \sin\left({\phi_0(t)}/{f}\right)$,  which is the same as in single field axion monodromy with
\be \label{dPhi1}
\dot\phi_1 = \frac{3b_*}{\alpha} \dot\phi_0 \cos\(\frac{\phi_0(t)}{f}\)~.
\ee 
This is the (small) oscillating part of the axion field velocity in addition to its constant slow rolling.
The equation of $\rho_1$ becomes
\be \label{rho1eq}
\ddot\rho_1 + 3 H\dot\rho_1 + \(m^2 -\frac{1}{2}\frac{\dot\phi_0^2}{\Lambda^2}\)\rho_1  = \frac{\dot\phi_0}{\Lambda} \dot\phi_1 \,.
\ee
The ${\dot\phi_0^2}/{\Lambda^2}$ term is subleading compared to the mass as required by the moduli stability. With the solution of $\phi_1$ in \eqref{dPhi1}, the source term on the right-hand side plays the role as an {\it oscillating driving force}. 
The solution of \eqref{rho1eq} contains two parts: the homogeneous solution corresponds to the oscillatory decay of a heavy scalar in de Sitter spacetime;  the particular solution captures the periodic modulation by the axion. Neglecting the damped heavy field oscillation, we  find the following result
\be \label{rho1}
\rho_1 = B \cos \left(\frac{\phi_0(t)}{f}+\delta \right)~,
\ee 
where $B= -{A^4}/(\Lambda\Xi^2)$ and $\delta = \arcsin{\left(({m^2-\omega^2})/\Xi^2\right)} $ with
$\Xi^4=9\omega^2 H^2 + (\omega^2-m^2)^2$. To estimate the relative size of oscillations in the $\rho$ direction, we find $B \simeq  b_* f^2/\Lambda$ in the regime $\omega\gtrsim m\gg H$. {For $\omega\gg m$, $B\simeq 3b_*f^2/(\alpha \Lambda)$ is more suppressed. Meanwhile, we notice that the size of the $\rho$ oscillation depends on the specific model. For instance, our choice here neglects the coupling between axion and modulus field in potential, which can lead to further enhancement of $B$. In the rest of the paper, we will show that the size of $B$ directly determines the magnitude of mixing between curvature and isocurvature fields. Therefore, our results can be understood as a minimum estimation.}

In summary, through controlled computation, we derived the background evolution of the two-field model \eqref{theory}, which displays oscillations in both the axion and modulus field directions with the same frequency $\omega$.\footnote{Note that this type of trajectories with constant wiggles differ from the ones with damped oscillations, which are normally generated by sharp turns and heavy masses  \cite{Chen:2011zf,Shiu:2011qw,Gao:2012uq,Chen:2014cwa}.}
For illustration, we plot one curved trajectory with wiggles in Figure \ref{fig:axion}. 
The intuitive explanation goes as follows: when the axion velocity acquires small oscillations on the top of its slow-roll motion, the centrifugal force due to the kinetic mixing drives oscillatory deviations from the stabilised position of the heavy field. 
This background behaviour is generally expected in axion monodromy models with periodically modulated potentials {(see \cite{Bhattacharya:2022fze,Chakraborty:2019dfh} for examples of rapid/sharp-turn trajectories with heavy fields)}.  

\vskip6pt
{\it Subtleties of mixings}-- 
For inflationary fluctuations, nontrivial consequences are expected for the wiggly trajectories.
At first thought, 
interactions of fluctuations in this two-field system can be directly read off from the original Lagrangian \eqref{theory}. By expanding the kinetic function $\exp{(\rho/\Lambda)}=1+\rho/\Lambda +...$, we see the mixings come from a dimension-five operator $\rho(\partial \phi)^2/\Lambda$. In flat gauge $\phi = \phi_B+ \delta\phi$ and $\rho=\rho_B+\delta\rho$, the two interactions are given by $(\dot\phi_B/\Lambda)\dot{\delta\phi} \delta\rho$ and $\delta\rho(\partial \delta\phi)^2/\Lambda$. Thus with the background solution $\dot\phi_1$, one can get oscillatory couplings for the linear mixing but not for the cubic vertex \cite{Chen:2022vzh}.
However, this simple consideration is not suited for computing the inflationary observable, i.e. primordial curvature perturbation $\zeta$. 
Normally we use a field redefinition to build the connection $\delta\phi=(\dot\phi_B/H) \zeta$. Here due to the oscillating piece of $\dot\phi_B$,  this simple relation misses the resonance effects of field interactions.
Another subtlety concerns the oscillating trajectory: inflation does not take place exclusively in the $\phi$ direction, and thus the fluctuations $\delta\phi$ do not directly determine $\zeta$.

In this work, we take a more cautious approach to these subtleties. 
As preparation, let's first introduce the covariant formalism of multi-field inflation  \cite{Achucarro:2010da,Gong:2011uw}. 
The kinetic term in \eqref{theory} corresponds to a 2D hyperbolic field space with the coordinate $\Phi^a = (\rho, \phi)$ and the metric $G_{ab}={\rm diag}\{1, \exp{(\rho/\Lambda)}\}$. 
For the background solution $\Phi^a_B(t)=(\rho_1,\phi_0+\phi_1)$,  we introduce a new basis  at each point of the trajectory by defining the tangent and normal unit vectors
$T^a \equiv \dot\Phi^a_B/\dot\Phi_t$ and $N^a\equiv \sqrt{\det G} \epsilon^{ab} T_b$,
where the total field velocity is given by $\dot\Phi_{\rm t}^2 \equiv\dot\rho_1^2 + \exp{(\rho_1/\Lambda)} (\dot\phi_0+\dot\phi_1)^2 $.
Then  the turning rate of the  trajectory  can be defined in a covariant way
\begin{align} 
\label{turning}
\Omega \equiv T_a D_t N^a  & \simeq   \frac{1}{2\Lambda}\dot\phi_0 + \frac{1}{2\Lambda}\dot\phi_1 - \frac{\ddot\rho_1}{2\dot\phi_0}  + ... \\
&{\simeq\frac{\dot{\phi}_0}{2\Lambda}\left\{1+\left(3\frac{b_*}{\alpha}+b_*\right)\cos\left(\frac{\phi_0(t)}{f}\right)+...\right\}}\nonumber
\end{align}
where in the last step we have used the background solution and applied the conditions in \eqref{conditions}.
This parameter describes the deviation from the geodesic motion in a curved manifold. The first term describes a constant turn given by the 0th order background, while the last two are the leading oscillating contributions.
{By using the background solutions, one can check that a small turning rate $\Omega\ll H$ is ensured as long as $\Lambda \gg \dot\phi_0 /H$. Thus the turning-rate correction to the isocurvature mass, which plays an important role in other multi-field models, is negligible here. (The effect of an oscillatory isocurvature mass on the cosmological collider signals has been systematically investigated in \cite{Jazayeri:2025vlv}.)}

To identify the leading interactions between the curvature and isocurvature perturbations, we adopt the EFT of inflation approach \cite{Cheung:2007st} and extend it to multi-field scenarios. 
The starting point is the unitary gauge where field fluctuations along the trajectory vanish, and thus the perturbed scalar field can be written as $\Phi^a(t,{\bf x})= \Phi^a_B(t) + \sigma N^a$. 
Furthermore, as we are interested in the resonance effects, which occur deep inside the horizon, we take the decoupling limit and neglect the mixing with gravity \cite{Behbahani:2011it}. This simplification allows us to focus on field interactions, and then in unitary gauge we find that the mixing mainly comes from the kinetic term. Specifically, using $\partial_\mu\Phi^a =\delta_\mu^0 \dot\Phi^a_B +  \partial_\mu ( \sigma N^a )$, we obtain the interaction operator linear in $\sigma$ as
\bea \label{eft}
-\frac{1}{2}  g^{\mu\nu} G_{ab} \partial_\mu \Phi^a \partial_\nu \Phi^b 
&\subset & - \dot\Phi_{t}  G_{ab} g^{0\mu} T^a  \partial_\mu (\sigma N^b) \nonumber\\
&\subset & \lambda (t) \delta g^{00} \sigma ~,
\eea
where $\lambda (t)=-\dot\Phi_t \Omega~{\simeq -\frac{\dot{\phi}_0^2}{2\Lambda}\left[1+b_*\cos\left(\frac{\phi_0(t)}{f}+\delta\right)\right]}$ and in the last step we have used the definition of the turning rate in \eqref{turning} and $T^aN_a=0$.
The EFT operator in  \eqref{eft} gives us the dominant mixing between adiabatic and isocurvature perturbations. 
Next, we perform the gauge transformation to bring back the Goldstone $\pi$ of time diffeomorphism breaking, $\delta g^{00} \rightarrow -2\dot\pi +(\partial_\mu\pi)^2$. 
Now we see that \textit{both} the linear mixing $-2\lambda\dot\pi\sigma$ \textit{and} the cubic interaction $\lambda (\partial_\mu\pi)^2\sigma$ acquire oscillatory couplings proportional to the turning rate $\Omega$.
Meanwhile, when we move to the Goldstone gauge, because of the strong time-dependence of  $\lambda(t+\pi)$, in the EFT another cubic vertex appears as $-2\dot\lambda \pi \dot\pi \sigma$. Considering that time derivatives on highly oscillating functions lead to large prefactors  ${\dot\lambda\sim \omega \lambda_{\mathrm{osc}} \sim\omega b_*\lambda}$, {this vertex is more significant than $\lambda(\partial_\mu\pi)^2\sigma$ for $\alpha b_* > 1$}. In terms of curvature perturbations $\zeta =H\pi$, the leading mixing interactions with the isocurvature mode $\sigma$ are
\be \label{mix}
{\mathcal{L}_{\rm mix} =(\Bar{g}+ g_2) \dot\zeta\sigma + g_3  \zeta\dot\zeta\sigma+(\tilde{g}+\tilde{g}_3)(\partial_\mu\zeta)^2\sigma~,}
\ee
where $\Bar{g}=\dot\phi_0^2/(H\Lambda)$ and the other couplings are given by
\begin{small}
\begin{align}
&g_2 \simeq   \Bar{g} b_* \cos \(\frac{\phi_0(t)}{f}+\delta \)  
,~g_3  \simeq  \Bar{g} \alpha b_* \sin \(\frac{\phi_0(t)}{f}+\delta \)  ~, \nonumber\\
&{\tilde{g} = \frac{\bar{g}}{2H},~~~~~~~~~~~~~~~~~~~~~~~~\tilde{g}_3 = \tilde{g}
b_*\cos\left(\frac{\phi_0(t)}{f}+\delta\right)}.
\label{coupling}
\end{align}
\end{small}
The cubic vertex ${g_3\simeq\bar{g}\alpha b_*\sin\(\frac{\phi_0(t)}{f}+\delta\)}$
can be seen as an analogy of $\epsilon\dot\eta\zeta^2\dot\zeta$ in single field axion monodromy, which gives the leading contribution for resonant non-Gaussianity \cite{Flauger:2010ja}. As a consistency check, we notice that \eqref{mix} agrees with the full result of quadratic and cubic actions of multi-field inflation \cite{Garcia-Saenz:2019njm} when we consider a highly oscillatory trajectory. {Also, taking the limit $b_*\rightarrow0$, all oscillatory coupling coefficients vanish. The model then smoothly reduces to the standard cosmological collider scenario without oscillatory features.}

\vskip6pt

{\it The moduli strike back}--
With the above knowledge, we briefly examine the validity of the single field effective description of axion monodromy. 
The wiggly trajectory threatens to generate interesting multi-field effects. Now we show under which conditions
we can no longer integrate out the modulus to achieve a single field EFT.

We follow the EFT approach of \cite{Achucarro:2010da, Baumann:2011su,Achucarro:2012sm} and focus on the regime of large moduli masses. The equation of motion of the heavy isocurvature field with the linear mixing $\dot\pi\sigma$ is given by
$ 
\ddot\sigma+ 3H\dot \sigma - \frac{1}{a^2}\partial_i^2 \sigma + m^2 \sigma =  2 \lambda \dot\pi .
$
When $k^2/a^2\ll m^2$, the isocurvature modes have the approximate solution $\sigma_0 = (2\lambda/m^2)\dot\pi$. Substituting this into the perturbation action, we find that the reduction of the Goldstone's sound speed is negligible $c_s^{-2} -1 = {\dot\Phi_t^2}/{(\Lambda m)^2}\rightarrow 0$ for sufficiently large $\Lambda$. Naively this would suggest that the single field description can be recovered and the moduli fields are decoupled.

However, the wiggly inflaton trajectory invalidates the analysis above.  In deriving the approximate solution $\sigma_0$, one underlying assumption is $\ddot\sigma\ll m^2\sigma$.\footnote{{Note that in principle the full isocurvature mass receives contributions from turning and field space curvature. These corrections are negligible in our setup, and the mass can be approximated by the Hessian of the potential $m^2=W''$.}} But for axion monodromy models we can check that $\ddot\sigma_0 \simeq \omega^2 \sigma_0$ due to the oscillatory coupling $\lambda(t)$. Thus the procedure of integrating out $\sigma$ field is valid only for $\omega \ll m$.\footnote{This corresponds to the adiabaticity condition proposed in \cite{Cespedes:2012hu, Achucarro:2012yr}, which is commonly used to examine the validity of single field  EFT for sharp-turn trajectories.}
In the parameter regime $\omega \gtrsim m$, the moduli fields get continuously excited by the background oscillations and a full treatment with multiple fields is required. 
As a result,  axion monodromy inflation becomes sensitive to UV physics much above the Hubble scale. It is worth noting that even in the most conservative scenario with $\Lambda = \mpl$, we still expect significantly large  $\ddot \lambda$ to render the single field EFT invalid.

\begin{figure} 
\centering
\includegraphics[width=0.7\linewidth]{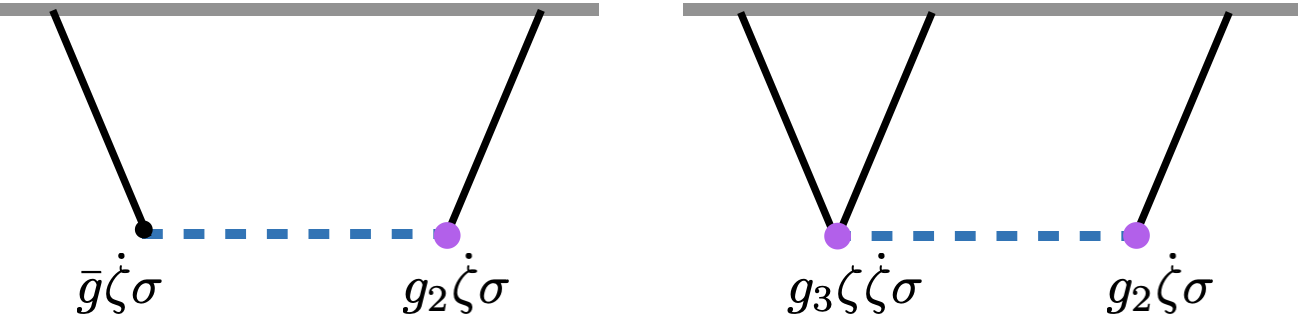}
\caption{The Feynman diagrams with leading resonance contributions to the $\zeta$ power spectrum and bispectrum. The purple dots denote vertices with oscillating couplings.}
\label{fig:feynman}
\end{figure}

\vskip6pt

{\it Cosmological collider, amplified}--
Now let's study new signatures of heavy moduli in cosmological correlators. 
We leave the detailed computation using the bootstrap method to \cite{ResonantCollider}, {where we follow the methodology in \cite{Pimentel:2022fsc}, and start with constructing the three-point scalar seed with two conformally coupled scalars, which can be fully solved by using the boundary differential equations. Taking it as the building block, the full bispectrum can be derived with weight-shifting operators corresponding to $\zeta\dot{\zeta}\sigma$ and $(\partial_{\mu}\zeta)^2\sigma$.} Here, we simply collect the final results with a focus on the phenomenology of non-Gaussianity. See Ref.  \cite{Chen:2022vzh, Qin:2023ejc, Werth:2023pfl, Pinol:2023oux} for related discussions on massive fields with oscillatory couplings during inflation.

In our setup, three types of oscillations are present for field fluctuations on sub-Hubble scales: the standard Bunch-Davies vacuum of the inflaton $\delta\phi_t \sim e^{ik\eta} $, the massive oscillations of the isocurvature mode $\sigma \sim e^{i m t}$, and the couplings $g_2,g_3 \sim \cos(\omega t + \delta)$.
In single field axion monodromy, sizable resonant non-Gaussianity is generated by the interplay between Bunch-Davies and oscillating couplings. 
For the two-field regime of axion monodromy, in the computation of non-Gaussianity, we encounter integrals of {the schematic form} 
\begin{small}
\be \label{resonance}
\int d\eta \frac{e^{i(k_1+k_2)\eta}}{\eta^{2+i\alpha}} \sigma^*_{k_3}(\eta) \sim e^{\pi(\alpha-\mu)/2}\Gamma\(\frac{1}{2}-i\alpha+i\mu\)~,
\ee
\end{small}with $\mu=\sqrt{m^2/H^2-9/4}\gg 1$.
When $\alpha<\mu$, this time integral leads to the familiar Boltzmann suppression factor $e^{-\pi \mu}$. In our regime of interest, $\alpha\gtrsim \mu\gg1$, the oscillatory coupling provides an extra resonance enhancement that overcomes the suppression effect,  {as expected from the general analysis of \cite{Chen:2022vzh}}. 

The price to pay is that we break scale invariance. Thus we also expect oscillations in the primordial power spectrum $P_\zeta = P_0[1+\delta n \cos(\omega \log(k/k_*))]$ with $P_0=H^2/(4 \epsilon_0\mpl^2 k^3) $ being the featureless component. From the left-hand Feynman diagram in Figure \ref{fig:feynman} we find 
\begin{align}
    \delta n^{\mathrm{col.}} \simeq -2\epsilon_0\frac{M_{\mathrm{Pl}}^2}{\Lambda^2}|E_1^P(\mu,\alpha)|b_*~,
    \label{4.3E4}
\end{align}
where $E_1^P(\mu,\alpha)$ is a known prefactor of $\mathcal{O}(0.01)$, {which we omit for brevity}. 
The single field results contain the same type of correction with $\delta n^{\rm s.f.} = 3b_*\sqrt{2\pi /\alpha}$ \cite{Flauger:2009ab}. Thus with small $\Lambda$ and/or large $\alpha$, \eqref{4.3E4} can be more dominant.
Meanwhile, the Planck constraint on this type of correction is  $\delta n \lesssim 0.05$ \cite{Planck:2018jri}. 

\begin{figure} 
\centering
\includegraphics[width=0.99\linewidth]{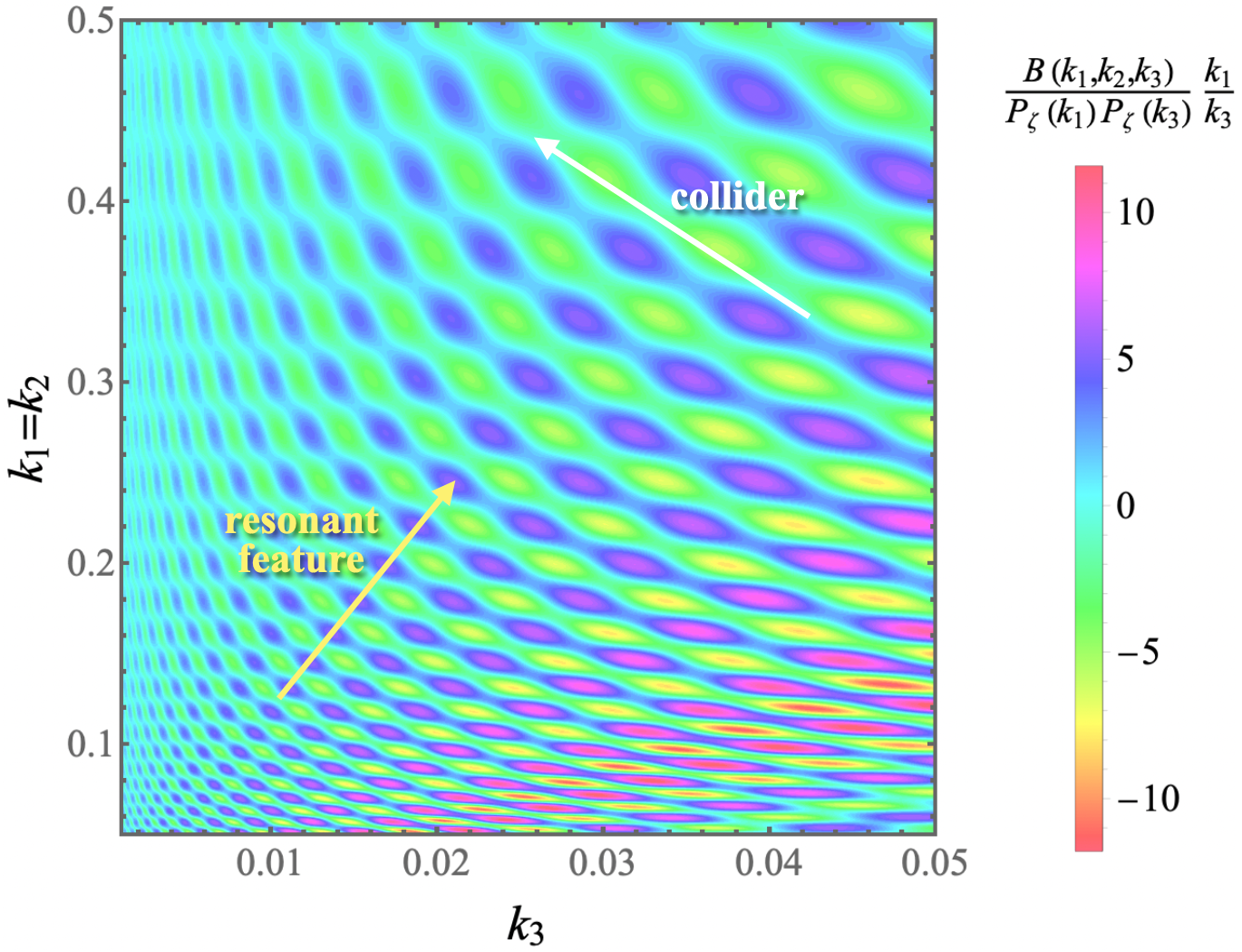}
\caption{Oscillatory pattern in the squeezed  bispectrum for $\alpha = 30$ and $\mu=10$ (which has $E_1^P=0.016$ and $E_1^B=0.128$). The shape has both {the scale-invariant} collider signals {perpendicular to the $k_3/k_1={\rm const.}$ lines} and scale-dependent oscillations along the $k_3/k_1={\rm const.}$ lines, {which is significantly distinct from standard resonant/collider templates.}}
\label{fig:sq}
\end{figure}

For the bispectrum, the dominant contribution corresponds to the case that both the quadratic and the cubic vertices oscillate (see the right-hand Feynman diagram in Figure \ref{fig:feynman}). 
In the companion paper \cite{ResonantCollider} we derived the full shape $\langle \zeta_{{\bf k}_1} \zeta_{{\bf k}_2} \zeta_{{\bf k}_3} \rangle = B(k_1,k_2,k_3)(2\pi)^3\delta^{(3)}({\bf k}_1+{\bf k}_2+{\bf k}_3)$  using the boundary differential equation of the bootstrap method. In this Letter, we focus on the squeezed limit where the resonant cosmological collider is manifest 
\begin{small}
\begin{align}
&\lim_{k_3\ll k_1}  B(k_1,k_2,k_3)  = \fnl^{\rm col.} P_{\zeta}(k_1)P_{\zeta}(k_3)
     \left(\frac{k_3}{k_1}\right)^{3/2}
     \label{bispectrum}\\
     & ~~~~~~~~~~~~\times \left\{\cos\left[(\alpha+\mu)\log\left(\frac{k_3}{k_1}\right)-2\alpha\log\left(k_3\right)+\delta_x\right]\right.\nonumber\\
     & ~~~~~~~~~~~~~~~~~\left.+\Delta f\cos\left[(\alpha-\mu)\log\left(\frac{k_3}{k_1}\right)+\delta_y\right]\right\}~,\nonumber
\end{align}
\end{small}
where $\Delta f$ is a dimensionless factor of $\mathcal{O}(1)$, {which only appear if the cubic interaction is oscillatory}. We find a distinctive signature with both the resonant-type scale-dependent non-Gaussianity and also the enhanced collider signal with heavy masses. The oscillatory pattern is shown in Fig. \ref{fig:sq}. 
The size of the signal is given by
\begin{align}
    f_{\mathrm{NL}}^{\mathrm{col.}} \simeq -\frac{1}{2}\epsilon_0\frac{M_{\mathrm{Pl}}^2}{\Lambda^2}|E_1^B(\mu,\alpha)|\alpha^2 b_*^2 ~,
    \label{4.3E5}
\end{align}
where $E_1^B$ from the bulk time integration is a combination of Gamma functions and hypergeometrics depending on $\mu$ and $\alpha$. In the featureless case with $\alpha=0$, this prefactor gives the Boltzmann suppression $E_1^B \sim e^{-\pi\mu}$.
But for $\alpha\gtrsim \mu\gg1$ we find $E_1^B \sim 0.1$ due to resonance enhancement discussed in \eqref{resonance}. {The main effects of oscillatory couplings on the cosmological collider signal can be summarised as follows
\begin{itemize}
    \item The oscillation in the linear mixing plays the dominant role in resonantly enhancing the cosmological collider signal. If the linear mixing term is non-oscillatory, the overall Boltzmann suppression remains.
    \item If the linear mixing oscillates while the cubic mixing does not, the leading term in full solution overcomes Boltzmann suppression. This give rise to the first term in (\ref{bispectrum}), consistent with the findings of [\textcolor{blue}{19}].
    \item If both linear and cubic vertices oscillate, an additional unsuppressed conbribution appears (the second term in (\ref{bispectrum})), which further enhances the resulting bispectrum and leads to a more intricate oscillatory pattern. The model of this work falls in this category.
\end{itemize}}

Now let's estimate the size of non-Gaussianity. 
There are two effects that may amplify the cosmological collider signals.
The first one is the resonance that overcomes the Boltzmann suppression and is universal for $\alpha\gtrsim \mu$.
The second is the strength of couplings in \eqref{mix}, which is model-dependent.
For $b_*\rightarrow 0$ and $\Lambda\sim\mpl$, the couplings are generally weak and we get small $\fnl$. 
To achieve large collider signals, here we consider $b_*\sim 0.1$ and roughly set $b_*|{E_1^B}/{E_1^P}|=1$, then $\fnl^{\mathrm{col.}} \simeq   \delta n^{\rm col.}  \alpha^2 /4$. 
Thus if we lower $\Lambda$ to saturate the observational bound on $\delta n$,  the resonant collider signal can become larger than the single field prediction $\fnl^{\rm s.f.} = {3\sqrt{2\pi}}\alpha^{\frac{3}{2}}b_*/{8}= \alpha^2\delta n^{\rm s.f.}/8$ \cite{Flauger:2010ja}.
Meanwhile, an EFT bound requires $\alpha \ll 400$ \cite{Behbahani:2011it}, within which one can parametrically achieve $\fnl^{\rm col.}$ of $\mathcal{O}(100)$.

{We emphasise that our computation relies on the assumption 
$B \simeq b_* f^2 / \Lambda$, which is valid when $\omega \gtrsim m$.
In the regime $\omega \gg m$, however, this assumption breaks down, and
the modulus oscillation is further suppressed by a factor of $H / \omega$.
As a consequence, the oscillatory cubic couplings in
Eq.~(\ref{coupling}) are modified to
\begin{small}
\begin{equation}
g_3 \simeq 4 \bar{g} \, b_* \sin\!\left( \frac{\phi_0(t)}{f} + \delta \right),~
\tilde{g}_3 \simeq \tilde{g} \, \frac{b_*}{\alpha}
\cos\!\left( \frac{\phi_0(t)}{f} + \delta \right),
\end{equation}
\end{small}
while the constant cubic interaction remains unchanged. It is then evident
that the $\tilde{g}_3$ term becomes negligible. In this regime, the
relative contributions to the final bispectrum from the oscillatory
interaction  $g_3 \, \zeta \dot{\zeta} \sigma$ and the constant interaction
$\tilde{g} (\partial_\mu \zeta)^2 \sigma$ depend sensitively on the
magnitude of $b_*$.
If $b_*=\mathcal{O}(0.01)$, the contribution from the oscillatory cubic
term is suppressed, and the system effectively reduces to the scenario
studied in Ref.~\cite{Chen:2022vzh}. On the other hand, if $b_*$ is
modestly enhanced to $\mathcal{O}(0.1)$, the oscillatory and
non-oscillatory contributions can become comparable, leading to a more
intricate phenomenology. Since this regime obscures the underlying physical
interpretation, we do not pursue it further in this work.}
\vskip6pt

{\it Concluding remarks}-- Heavy moduli are generally expected in UV completions of inflation and they couple to axions through the kinetic term.
We investigate the two-field regime of axion monodromy for both background and perturbations, and identified a novel type of UV sensitivity. Remarkably, due to the periodic modulation of the axion potential, heavy moduli are continuously excited by the oscillating background, {realizing the mechanism of \cite{Chen:2022vzh} in a concrete and well-motivated model}. When the oscillation frequency becomes larger than the lightest moduli mass, this phenomenon leads to the breakdown of the effective single field description. Furthermore, we find a new type of unsuppressed cosmological collider signals with heavy masses. 

This concrete example from a {{string-inspired}} setup points out an exciting direction to probe new physics much heavier than the Hubble scale during inflation, {as anticipated in \cite{Chen:2022vzh}}. On the theory side, we expect implications on both cosmological correlators and string inflation. While we take a field-theoretic approach here assuming a 4D EFT from string compactifications, it would be interesting to consider a full 10D picture and examine its UV sensitivity. For instance, the moduli have physical meanings in the stringy description, such as the volume of the compactified dimensions, the location of D-branes, etc. How is this geometrical information imprinted in late-time correlators? We leave this question for future work. 

Meanwhile, the new phenomenology deserves a closer look. The scalar bispectrum here can be seen as a combination of resonant non-Gaussianity and cosmological collider, which contains rich oscillatory structure in the squeezed limit and can potentially be large. With current tools, we would be able to search for this new type of non-Gaussianity signals in the Planck data as shown in \cite{Sohn:2023fte,Sohn:2024xzd,Suman:2025vuf,Suman:2025tpv}. Certainly, our signal serves as an interesting target for upcoming surveys such as Simons Observatory and SphereX.

\vskip12pt

{\it Acknowledgements--}
We would like to thank
Ana Ach\'ucarro,
 Carlos Duaso Pueyo,  Zhehan Qin, Fernando Quevedo, Paul Shellard, Xi Tong, Gonzalo Villa  for helpful discussions.
DGW is partially supported by a Rubicon Postdoctoral Fellowship from the Netherlands Organisation for Scientific Research (NWO). BZ is supported by the Science and Technology Facilities Council (STFC) studentship. EP is supported by STFC consolidated grant ST/T000694/1 and ST/X000664/1 and by the EPSRC New Horizon grant EP/V017268/1.

\bibliography{references}

\begin{thebibliography}{57}%
\makeatletter
\providecommand \@ifxundefined [1]{%
 \@ifx{#1\undefined}
}%
\providecommand \@ifnum [1]{%
 \ifnum #1\expandafter \@firstoftwo
 \else \expandafter \@secondoftwo
 \fi
}%
\providecommand \@ifx [1]{%
 \ifx #1\expandafter \@firstoftwo
 \else \expandafter \@secondoftwo
 \fi
}%
\providecommand \natexlab [1]{#1}%
\providecommand \enquote  [1]{``#1''}%
\providecommand \bibnamefont  [1]{#1}%
\providecommand \bibfnamefont [1]{#1}%
\providecommand \citenamefont [1]{#1}%
\providecommand \href@noop [0]{\@secondoftwo}%
\providecommand \href [0]{\begingroup \@sanitize@url \@href}%
\providecommand \@href[1]{\@@startlink{#1}\@@href}%
\providecommand \@@href[1]{\endgroup#1\@@endlink}%
\providecommand \@sanitize@url [0]{\catcode `\\12\catcode `\$12\catcode
  `\&12\catcode `\#12\catcode `\^12\catcode `\_12\catcode `\%12\relax}%
\providecommand \@@startlink[1]{}%
\providecommand \@@endlink[0]{}%
\providecommand \url  [0]{\begingroup\@sanitize@url \@url }%
\providecommand \@url [1]{\endgroup\@href {#1}{\urlprefix }}%
\providecommand \urlprefix  [0]{URL }%
\providecommand \Eprint [0]{\href }%
\providecommand \doibase [0]{http://dx.doi.org/}%
\providecommand \selectlanguage [0]{\@gobble}%
\providecommand \bibinfo  [0]{\@secondoftwo}%
\providecommand \bibfield  [0]{\@secondoftwo}%
\providecommand \translation [1]{[#1]}%
\providecommand \BibitemOpen [0]{}%
\providecommand \bibitemStop [0]{}%
\providecommand \bibitemNoStop [0]{.\EOS\space}%
\providecommand \EOS [0]{\spacefactor3000\relax}%
\providecommand \BibitemShut  [1]{\csname bibitem#1\endcsname}%
\let\auto@bib@innerbib\@empty
\bibitem [{\citenamefont {Baumann}\ and\ \citenamefont
  {McAllister}(2015)}]{Baumann:2014nda}%
  \BibitemOpen
  \bibfield  {author} {\bibinfo {author} {\bibfnamefont {D.}~\bibnamefont
  {Baumann}}\ and\ \bibinfo {author} {\bibfnamefont {L.}~\bibnamefont
  {McAllister}},\ }\href {\doibase 10.1017/CBO9781316105733} {\emph {\bibinfo
  {title} {{Inflation and String Theory}}}},\ Cambridge Monographs on
  Mathematical Physics\ (\bibinfo  {publisher} {Cambridge University Press},\
  \bibinfo {year} {2015})\ \Eprint {http://arxiv.org/abs/1404.2601}
  {arXiv:1404.2601 [hep-th]} \BibitemShut {NoStop}%
\bibitem [{\citenamefont {Cicoli}\ \emph {et~al.}(2024)\citenamefont {Cicoli},
  \citenamefont {Conlon}, \citenamefont {Maharana}, \citenamefont
  {Parameswaran}, \citenamefont {Quevedo},\ and\ \citenamefont
  {Zavala}}]{Cicoli:2023opf}%
  \BibitemOpen
  \bibfield  {author} {\bibinfo {author} {\bibfnamefont {M.}~\bibnamefont
  {Cicoli}}, \bibinfo {author} {\bibfnamefont {J.~P.}\ \bibnamefont {Conlon}},
  \bibinfo {author} {\bibfnamefont {A.}~\bibnamefont {Maharana}}, \bibinfo
  {author} {\bibfnamefont {S.}~\bibnamefont {Parameswaran}}, \bibinfo {author}
  {\bibfnamefont {F.}~\bibnamefont {Quevedo}}, \ and\ \bibinfo {author}
  {\bibfnamefont {I.}~\bibnamefont {Zavala}},\ }\href {\doibase
  10.1016/j.physrep.2024.01.002} {\bibfield  {journal} {\bibinfo  {journal}
  {Phys. Rept.}\ }\textbf {\bibinfo {volume} {1059}},\ \bibinfo {pages} {1}
  (\bibinfo {year} {2024})},\ \Eprint {http://arxiv.org/abs/2303.04819}
  {arXiv:2303.04819 [hep-th]} \BibitemShut {NoStop}%
\bibitem [{\citenamefont {Tolley}\ and\ \citenamefont
  {Wyman}(2010)}]{Tolley:2009fg}%
  \BibitemOpen
  \bibfield  {author} {\bibinfo {author} {\bibfnamefont {A.~J.}\ \bibnamefont
  {Tolley}}\ and\ \bibinfo {author} {\bibfnamefont {M.}~\bibnamefont {Wyman}},\
  }\href {\doibase 10.1103/PhysRevD.81.043502} {\bibfield  {journal} {\bibinfo
  {journal} {Phys. Rev. D}\ }\textbf {\bibinfo {volume} {81}},\ \bibinfo
  {pages} {043502} (\bibinfo {year} {2010})},\ \Eprint
  {http://arxiv.org/abs/0910.1853} {arXiv:0910.1853 [hep-th]} \BibitemShut
  {NoStop}%
\bibitem [{\citenamefont {Achucarro}\ \emph {et~al.}(2011)\citenamefont
  {Achucarro}, \citenamefont {Gong}, \citenamefont {Hardeman}, \citenamefont
  {Palma},\ and\ \citenamefont {Patil}}]{Achucarro:2010da}%
  \BibitemOpen
  \bibfield  {author} {\bibinfo {author} {\bibfnamefont {A.}~\bibnamefont
  {Achucarro}}, \bibinfo {author} {\bibfnamefont {J.-O.}\ \bibnamefont {Gong}},
  \bibinfo {author} {\bibfnamefont {S.}~\bibnamefont {Hardeman}}, \bibinfo
  {author} {\bibfnamefont {G.~A.}\ \bibnamefont {Palma}}, \ and\ \bibinfo
  {author} {\bibfnamefont {S.~P.}\ \bibnamefont {Patil}},\ }\href {\doibase
  10.1088/1475-7516/2011/01/030} {\bibfield  {journal} {\bibinfo  {journal}
  {JCAP}\ }\textbf {\bibinfo {volume} {01}},\ \bibinfo {pages} {030} (\bibinfo
  {year} {2011})},\ \Eprint {http://arxiv.org/abs/1010.3693} {arXiv:1010.3693
  [hep-ph]} \BibitemShut {NoStop}%
\bibitem [{\citenamefont {Baumann}\ and\ \citenamefont
  {Green}(2011)}]{Baumann:2011su}%
  \BibitemOpen
  \bibfield  {author} {\bibinfo {author} {\bibfnamefont {D.}~\bibnamefont
  {Baumann}}\ and\ \bibinfo {author} {\bibfnamefont {D.}~\bibnamefont
  {Green}},\ }\href {\doibase 10.1088/1475-7516/2011/09/014} {\bibfield
  {journal} {\bibinfo  {journal} {JCAP}\ }\textbf {\bibinfo {volume} {09}},\
  \bibinfo {pages} {014} (\bibinfo {year} {2011})},\ \Eprint
  {http://arxiv.org/abs/1102.5343} {arXiv:1102.5343 [hep-th]} \BibitemShut
  {NoStop}%
\bibitem [{\citenamefont {Achucarro}\ \emph
  {et~al.}(2012{\natexlab{a}})\citenamefont {Achucarro}, \citenamefont {Gong},
  \citenamefont {Hardeman}, \citenamefont {Palma},\ and\ \citenamefont
  {Patil}}]{Achucarro:2012sm}%
  \BibitemOpen
  \bibfield  {author} {\bibinfo {author} {\bibfnamefont {A.}~\bibnamefont
  {Achucarro}}, \bibinfo {author} {\bibfnamefont {J.-O.}\ \bibnamefont {Gong}},
  \bibinfo {author} {\bibfnamefont {S.}~\bibnamefont {Hardeman}}, \bibinfo
  {author} {\bibfnamefont {G.~A.}\ \bibnamefont {Palma}}, \ and\ \bibinfo
  {author} {\bibfnamefont {S.~P.}\ \bibnamefont {Patil}},\ }\href {\doibase
  10.1007/JHEP05(2012)066} {\bibfield  {journal} {\bibinfo  {journal} {JHEP}\
  }\textbf {\bibinfo {volume} {05}},\ \bibinfo {pages} {066} (\bibinfo {year}
  {2012}{\natexlab{a}})},\ \Eprint {http://arxiv.org/abs/1201.6342}
  {arXiv:1201.6342 [hep-th]} \BibitemShut {NoStop}%
\bibitem [{\citenamefont {Chen}\ and\ \citenamefont
  {Wang}(2010)}]{Chen:2009zp}%
  \BibitemOpen
  \bibfield  {author} {\bibinfo {author} {\bibfnamefont {X.}~\bibnamefont
  {Chen}}\ and\ \bibinfo {author} {\bibfnamefont {Y.}~\bibnamefont {Wang}},\
  }\href {\doibase 10.1088/1475-7516/2010/04/027} {\bibfield  {journal}
  {\bibinfo  {journal} {JCAP}\ }\textbf {\bibinfo {volume} {04}},\ \bibinfo
  {pages} {027} (\bibinfo {year} {2010})},\ \Eprint
  {http://arxiv.org/abs/0911.3380} {arXiv:0911.3380 [hep-th]} \BibitemShut
  {NoStop}%
\bibitem [{\citenamefont {Baumann}\ and\ \citenamefont
  {Green}(2012)}]{Baumann:2011nk}%
  \BibitemOpen
  \bibfield  {author} {\bibinfo {author} {\bibfnamefont {D.}~\bibnamefont
  {Baumann}}\ and\ \bibinfo {author} {\bibfnamefont {D.}~\bibnamefont
  {Green}},\ }\href {\doibase 10.1103/PhysRevD.85.103520} {\bibfield  {journal}
  {\bibinfo  {journal} {Phys. Rev. D}\ }\textbf {\bibinfo {volume} {85}},\
  \bibinfo {pages} {103520} (\bibinfo {year} {2012})},\ \Eprint
  {http://arxiv.org/abs/1109.0292} {arXiv:1109.0292 [hep-th]} \BibitemShut
  {NoStop}%
\bibitem [{\citenamefont {Noumi}\ \emph {et~al.}(2013)\citenamefont {Noumi},
  \citenamefont {Yamaguchi},\ and\ \citenamefont {Yokoyama}}]{Noumi:2012vr}%
  \BibitemOpen
  \bibfield  {author} {\bibinfo {author} {\bibfnamefont {T.}~\bibnamefont
  {Noumi}}, \bibinfo {author} {\bibfnamefont {M.}~\bibnamefont {Yamaguchi}}, \
  and\ \bibinfo {author} {\bibfnamefont {D.}~\bibnamefont {Yokoyama}},\ }\href
  {\doibase 10.1007/JHEP06(2013)051} {\bibfield  {journal} {\bibinfo  {journal}
  {JHEP}\ }\textbf {\bibinfo {volume} {06}},\ \bibinfo {pages} {051} (\bibinfo
  {year} {2013})},\ \Eprint {http://arxiv.org/abs/1211.1624} {arXiv:1211.1624
  [hep-th]} \BibitemShut {NoStop}%
\bibitem [{\citenamefont {Arkani-Hamed}\ and\ \citenamefont
  {Maldacena}(2015)}]{Arkani-Hamed:2015bza}%
  \BibitemOpen
  \bibfield  {author} {\bibinfo {author} {\bibfnamefont {N.}~\bibnamefont
  {Arkani-Hamed}}\ and\ \bibinfo {author} {\bibfnamefont {J.}~\bibnamefont
  {Maldacena}},\ }\href@noop {} {\  (\bibinfo {year} {2015})},\ \Eprint
  {http://arxiv.org/abs/1503.08043} {arXiv:1503.08043 [hep-th]} \BibitemShut
  {NoStop}%
\bibitem [{\citenamefont {Wang}\ and\ \citenamefont
  {Xianyu}(2020)}]{Wang:2019gbi}%
  \BibitemOpen
  \bibfield  {author} {\bibinfo {author} {\bibfnamefont {L.-T.}\ \bibnamefont
  {Wang}}\ and\ \bibinfo {author} {\bibfnamefont {Z.-Z.}\ \bibnamefont
  {Xianyu}},\ }\href {\doibase 10.1007/JHEP02(2020)044} {\bibfield  {journal}
  {\bibinfo  {journal} {JHEP}\ }\textbf {\bibinfo {volume} {02}},\ \bibinfo
  {pages} {044} (\bibinfo {year} {2020})},\ \Eprint
  {http://arxiv.org/abs/1910.12876} {arXiv:1910.12876 [hep-ph]} \BibitemShut
  {NoStop}%
\bibitem [{\citenamefont {Bodas}\ \emph {et~al.}(2021)\citenamefont {Bodas},
  \citenamefont {Kumar},\ and\ \citenamefont {Sundrum}}]{Bodas:2020yho}%
  \BibitemOpen
  \bibfield  {author} {\bibinfo {author} {\bibfnamefont {A.}~\bibnamefont
  {Bodas}}, \bibinfo {author} {\bibfnamefont {S.}~\bibnamefont {Kumar}}, \ and\
  \bibinfo {author} {\bibfnamefont {R.}~\bibnamefont {Sundrum}},\ }\href
  {\doibase 10.1007/JHEP02(2021)079} {\bibfield  {journal} {\bibinfo  {journal}
  {JHEP}\ }\textbf {\bibinfo {volume} {02}},\ \bibinfo {pages} {079} (\bibinfo
  {year} {2021})},\ \Eprint {http://arxiv.org/abs/2010.04727} {arXiv:2010.04727
  [hep-ph]} \BibitemShut {NoStop}%
\bibitem [{\citenamefont {Tong}\ and\ \citenamefont
  {Xianyu}(2022)}]{Tong:2022cdz}%
  \BibitemOpen
  \bibfield  {author} {\bibinfo {author} {\bibfnamefont {X.}~\bibnamefont
  {Tong}}\ and\ \bibinfo {author} {\bibfnamefont {Z.-Z.}\ \bibnamefont
  {Xianyu}},\ }\href@noop {} {\  (\bibinfo {year} {2022})},\ \Eprint
  {http://arxiv.org/abs/2203.06349} {arXiv:2203.06349 [hep-ph]} \BibitemShut
  {NoStop}%
\bibitem [{\citenamefont {Lee}\ \emph {et~al.}(2016)\citenamefont {Lee},
  \citenamefont {Baumann},\ and\ \citenamefont {Pimentel}}]{Lee:2016vti}%
  \BibitemOpen
  \bibfield  {author} {\bibinfo {author} {\bibfnamefont {H.}~\bibnamefont
  {Lee}}, \bibinfo {author} {\bibfnamefont {D.}~\bibnamefont {Baumann}}, \ and\
  \bibinfo {author} {\bibfnamefont {G.~L.}\ \bibnamefont {Pimentel}},\ }\href
  {\doibase 10.1007/JHEP12(2016)040} {\bibfield  {journal} {\bibinfo  {journal}
  {JHEP}\ }\textbf {\bibinfo {volume} {12}},\ \bibinfo {pages} {040} (\bibinfo
  {year} {2016})},\ \Eprint {http://arxiv.org/abs/1607.03735} {arXiv:1607.03735
  [hep-th]} \BibitemShut {NoStop}%
\bibitem [{\citenamefont {Pimentel}\ and\ \citenamefont
  {Wang}(2022)}]{Pimentel:2022fsc}%
  \BibitemOpen
  \bibfield  {author} {\bibinfo {author} {\bibfnamefont {G.~L.}\ \bibnamefont
  {Pimentel}}\ and\ \bibinfo {author} {\bibfnamefont {D.-G.}\ \bibnamefont
  {Wang}},\ }\href {\doibase 10.1007/JHEP10(2022)177} {\bibfield  {journal}
  {\bibinfo  {journal} {JHEP}\ }\textbf {\bibinfo {volume} {10}},\ \bibinfo
  {pages} {177} (\bibinfo {year} {2022})},\ \Eprint
  {http://arxiv.org/abs/2205.00013} {arXiv:2205.00013 [hep-th]} \BibitemShut
  {NoStop}%
\bibitem [{\citenamefont {Jazayeri}\ and\ \citenamefont
  {Renaux-Petel}(2022)}]{Jazayeri:2022kjy}%
  \BibitemOpen
  \bibfield  {author} {\bibinfo {author} {\bibfnamefont {S.}~\bibnamefont
  {Jazayeri}}\ and\ \bibinfo {author} {\bibfnamefont {S.}~\bibnamefont
  {Renaux-Petel}},\ }\href@noop {} {\  (\bibinfo {year} {2022})},\ \Eprint
  {http://arxiv.org/abs/2205.10340} {arXiv:2205.10340 [hep-th]} \BibitemShut
  {NoStop}%
\bibitem [{\citenamefont {Wang}\ \emph {et~al.}(2023)\citenamefont {Wang},
  \citenamefont {Pimentel},\ and\ \citenamefont {Ach\'ucarro}}]{Wang:2022eop}%
  \BibitemOpen
  \bibfield  {author} {\bibinfo {author} {\bibfnamefont {D.-G.}\ \bibnamefont
  {Wang}}, \bibinfo {author} {\bibfnamefont {G.~L.}\ \bibnamefont {Pimentel}},
  \ and\ \bibinfo {author} {\bibfnamefont {A.}~\bibnamefont {Ach\'ucarro}},\
  }\href {\doibase 10.1088/1475-7516/2023/05/043} {\bibfield  {journal}
  {\bibinfo  {journal} {JCAP}\ }\textbf {\bibinfo {volume} {05}},\ \bibinfo
  {pages} {043} (\bibinfo {year} {2023})},\ \Eprint
  {http://arxiv.org/abs/2212.14035} {arXiv:2212.14035 [astro-ph.CO]}
  \BibitemShut {NoStop}%
\bibitem [{\citenamefont {Jazayeri}\ \emph {et~al.}(2023)\citenamefont
  {Jazayeri}, \citenamefont {Renaux-Petel}, \citenamefont {Tong}, \citenamefont
  {Werth},\ and\ \citenamefont {Zhu}}]{Jazayeri:2023kji}%
  \BibitemOpen
  \bibfield  {author} {\bibinfo {author} {\bibfnamefont {S.}~\bibnamefont
  {Jazayeri}}, \bibinfo {author} {\bibfnamefont {S.}~\bibnamefont
  {Renaux-Petel}}, \bibinfo {author} {\bibfnamefont {X.}~\bibnamefont {Tong}},
  \bibinfo {author} {\bibfnamefont {D.}~\bibnamefont {Werth}}, \ and\ \bibinfo
  {author} {\bibfnamefont {Y.}~\bibnamefont {Zhu}},\ }\href {\doibase
  10.1103/PhysRevD.108.123523} {\bibfield  {journal} {\bibinfo  {journal}
  {Phys. Rev. D}\ }\textbf {\bibinfo {volume} {108}},\ \bibinfo {pages}
  {123523} (\bibinfo {year} {2023})},\ \Eprint
  {http://arxiv.org/abs/2308.11315} {arXiv:2308.11315 [hep-th]} \BibitemShut
  {NoStop}%
\bibitem [{\citenamefont {Chen}\ \emph {et~al.}(2022)\citenamefont {Chen},
  \citenamefont {Ebadi},\ and\ \citenamefont {Kumar}}]{Chen:2022vzh}%
  \BibitemOpen
  \bibfield  {author} {\bibinfo {author} {\bibfnamefont {X.}~\bibnamefont
  {Chen}}, \bibinfo {author} {\bibfnamefont {R.}~\bibnamefont {Ebadi}}, \ and\
  \bibinfo {author} {\bibfnamefont {S.}~\bibnamefont {Kumar}},\ }\href
  {\doibase 10.1088/1475-7516/2022/08/083} {\bibfield  {journal} {\bibinfo
  {journal} {JCAP}\ }\textbf {\bibinfo {volume} {08}},\ \bibinfo {pages} {083}
  (\bibinfo {year} {2022})},\ \Eprint {http://arxiv.org/abs/2205.01107}
  {arXiv:2205.01107 [hep-ph]} \BibitemShut {NoStop}%
\bibitem [{\citenamefont {Silverstein}\ and\ \citenamefont
  {Westphal}(2008)}]{Silverstein:2008sg}%
  \BibitemOpen
  \bibfield  {author} {\bibinfo {author} {\bibfnamefont {E.}~\bibnamefont
  {Silverstein}}\ and\ \bibinfo {author} {\bibfnamefont {A.}~\bibnamefont
  {Westphal}},\ }\href {\doibase 10.1103/PhysRevD.78.106003} {\bibfield
  {journal} {\bibinfo  {journal} {Phys. Rev. D}\ }\textbf {\bibinfo {volume}
  {78}},\ \bibinfo {pages} {106003} (\bibinfo {year} {2008})},\ \Eprint
  {http://arxiv.org/abs/0803.3085} {arXiv:0803.3085 [hep-th]} \BibitemShut
  {NoStop}%
\bibitem [{\citenamefont {McAllister}\ \emph {et~al.}(2010)\citenamefont
  {McAllister}, \citenamefont {Silverstein},\ and\ \citenamefont
  {Westphal}}]{McAllister:2008hb}%
  \BibitemOpen
  \bibfield  {author} {\bibinfo {author} {\bibfnamefont {L.}~\bibnamefont
  {McAllister}}, \bibinfo {author} {\bibfnamefont {E.}~\bibnamefont
  {Silverstein}}, \ and\ \bibinfo {author} {\bibfnamefont {A.}~\bibnamefont
  {Westphal}},\ }\href {\doibase 10.1103/PhysRevD.82.046003} {\bibfield
  {journal} {\bibinfo  {journal} {Phys. Rev. D}\ }\textbf {\bibinfo {volume}
  {82}},\ \bibinfo {pages} {046003} (\bibinfo {year} {2010})},\ \Eprint
  {http://arxiv.org/abs/0808.0706} {arXiv:0808.0706 [hep-th]} \BibitemShut
  {NoStop}%
\bibitem [{\citenamefont {Flauger}\ \emph {et~al.}(2010)\citenamefont
  {Flauger}, \citenamefont {McAllister}, \citenamefont {Pajer}, \citenamefont
  {Westphal},\ and\ \citenamefont {Xu}}]{Flauger:2009ab}%
  \BibitemOpen
  \bibfield  {author} {\bibinfo {author} {\bibfnamefont {R.}~\bibnamefont
  {Flauger}}, \bibinfo {author} {\bibfnamefont {L.}~\bibnamefont {McAllister}},
  \bibinfo {author} {\bibfnamefont {E.}~\bibnamefont {Pajer}}, \bibinfo
  {author} {\bibfnamefont {A.}~\bibnamefont {Westphal}}, \ and\ \bibinfo
  {author} {\bibfnamefont {G.}~\bibnamefont {Xu}},\ }\href {\doibase
  10.1088/1475-7516/2010/06/009} {\bibfield  {journal} {\bibinfo  {journal}
  {JCAP}\ }\textbf {\bibinfo {volume} {06}},\ \bibinfo {pages} {009} (\bibinfo
  {year} {2010})},\ \Eprint {http://arxiv.org/abs/0907.2916} {arXiv:0907.2916
  [hep-th]} \BibitemShut {NoStop}%
\bibitem [{\citenamefont {Berg}\ \emph {et~al.}(2010)\citenamefont {Berg},
  \citenamefont {Pajer},\ and\ \citenamefont {Sjors}}]{Berg:2009tg}%
  \BibitemOpen
  \bibfield  {author} {\bibinfo {author} {\bibfnamefont {M.}~\bibnamefont
  {Berg}}, \bibinfo {author} {\bibfnamefont {E.}~\bibnamefont {Pajer}}, \ and\
  \bibinfo {author} {\bibfnamefont {S.}~\bibnamefont {Sjors}},\ }\href
  {\doibase 10.1103/PhysRevD.81.103535} {\bibfield  {journal} {\bibinfo
  {journal} {Phys. Rev. D}\ }\textbf {\bibinfo {volume} {81}},\ \bibinfo
  {pages} {103535} (\bibinfo {year} {2010})},\ \Eprint
  {http://arxiv.org/abs/0912.1341} {arXiv:0912.1341 [hep-th]} \BibitemShut
  {NoStop}%
\bibitem [{\citenamefont {Kaloper}\ \emph {et~al.}(2011)\citenamefont
  {Kaloper}, \citenamefont {Lawrence},\ and\ \citenamefont
  {Sorbo}}]{Kaloper:2011jz}%
  \BibitemOpen
  \bibfield  {author} {\bibinfo {author} {\bibfnamefont {N.}~\bibnamefont
  {Kaloper}}, \bibinfo {author} {\bibfnamefont {A.}~\bibnamefont {Lawrence}}, \
  and\ \bibinfo {author} {\bibfnamefont {L.}~\bibnamefont {Sorbo}},\ }\href
  {\doibase 10.1088/1475-7516/2011/03/023} {\bibfield  {journal} {\bibinfo
  {journal} {JCAP}\ }\textbf {\bibinfo {volume} {03}},\ \bibinfo {pages} {023}
  (\bibinfo {year} {2011})},\ \Eprint {http://arxiv.org/abs/1101.0026}
  {arXiv:1101.0026 [hep-th]} \BibitemShut {NoStop}%
\bibitem [{\citenamefont {Chen}\ \emph {et~al.}(2008)\citenamefont {Chen},
  \citenamefont {Easther},\ and\ \citenamefont {Lim}}]{Chen:2008wn}%
  \BibitemOpen
  \bibfield  {author} {\bibinfo {author} {\bibfnamefont {X.}~\bibnamefont
  {Chen}}, \bibinfo {author} {\bibfnamefont {R.}~\bibnamefont {Easther}}, \
  and\ \bibinfo {author} {\bibfnamefont {E.~A.}\ \bibnamefont {Lim}},\ }\href
  {\doibase 10.1088/1475-7516/2008/04/010} {\bibfield  {journal} {\bibinfo
  {journal} {JCAP}\ }\textbf {\bibinfo {volume} {04}},\ \bibinfo {pages} {010}
  (\bibinfo {year} {2008})},\ \Eprint {http://arxiv.org/abs/0801.3295}
  {arXiv:0801.3295 [astro-ph]} \BibitemShut {NoStop}%
\bibitem [{\citenamefont {Flauger}\ and\ \citenamefont
  {Pajer}(2011)}]{Flauger:2010ja}%
  \BibitemOpen
  \bibfield  {author} {\bibinfo {author} {\bibfnamefont {R.}~\bibnamefont
  {Flauger}}\ and\ \bibinfo {author} {\bibfnamefont {E.}~\bibnamefont
  {Pajer}},\ }\href {\doibase 10.1088/1475-7516/2011/01/017} {\bibfield
  {journal} {\bibinfo  {journal} {JCAP}\ }\textbf {\bibinfo {volume} {01}},\
  \bibinfo {pages} {017} (\bibinfo {year} {2011})},\ \Eprint
  {http://arxiv.org/abs/1002.0833} {arXiv:1002.0833 [hep-th]} \BibitemShut
  {NoStop}%
\bibitem [{\citenamefont {Chen}(2010)}]{Chen:2010bka}%
  \BibitemOpen
  \bibfield  {author} {\bibinfo {author} {\bibfnamefont {X.}~\bibnamefont
  {Chen}},\ }\href {\doibase 10.1088/1475-7516/2010/12/003} {\bibfield
  {journal} {\bibinfo  {journal} {JCAP}\ }\textbf {\bibinfo {volume} {12}},\
  \bibinfo {pages} {003} (\bibinfo {year} {2010})},\ \Eprint
  {http://arxiv.org/abs/1008.2485} {arXiv:1008.2485 [hep-th]} \BibitemShut
  {NoStop}%
\bibitem [{\citenamefont {Behbahani}\ \emph {et~al.}(2012)\citenamefont
  {Behbahani}, \citenamefont {Dymarsky}, \citenamefont {Mirbabayi},\ and\
  \citenamefont {Senatore}}]{Behbahani:2011it}%
  \BibitemOpen
  \bibfield  {author} {\bibinfo {author} {\bibfnamefont {S.~R.}\ \bibnamefont
  {Behbahani}}, \bibinfo {author} {\bibfnamefont {A.}~\bibnamefont {Dymarsky}},
  \bibinfo {author} {\bibfnamefont {M.}~\bibnamefont {Mirbabayi}}, \ and\
  \bibinfo {author} {\bibfnamefont {L.}~\bibnamefont {Senatore}},\ }\href
  {\doibase 10.1088/1475-7516/2012/12/036} {\bibfield  {journal} {\bibinfo
  {journal} {JCAP}\ }\textbf {\bibinfo {volume} {12}},\ \bibinfo {pages} {036}
  (\bibinfo {year} {2012})},\ \Eprint {http://arxiv.org/abs/1111.3373}
  {arXiv:1111.3373 [hep-th]} \BibitemShut {NoStop}%
\bibitem [{\citenamefont {Leblond}\ and\ \citenamefont
  {Pajer}(2011)}]{Leblond:2010yq}%
  \BibitemOpen
  \bibfield  {author} {\bibinfo {author} {\bibfnamefont {L.}~\bibnamefont
  {Leblond}}\ and\ \bibinfo {author} {\bibfnamefont {E.}~\bibnamefont
  {Pajer}},\ }\href {\doibase 10.1088/1475-7516/2011/01/035} {\bibfield
  {journal} {\bibinfo  {journal} {JCAP}\ }\textbf {\bibinfo {volume} {01}},\
  \bibinfo {pages} {035} (\bibinfo {year} {2011})},\ \Eprint
  {http://arxiv.org/abs/1010.4565} {arXiv:1010.4565 [hep-th]} \BibitemShut
  {NoStop}%
\bibitem [{\citenamefont {Cabass}\ \emph {et~al.}(2018)\citenamefont {Cabass},
  \citenamefont {Pajer},\ and\ \citenamefont {Schmidt}}]{Cabass:2018roz}%
  \BibitemOpen
  \bibfield  {author} {\bibinfo {author} {\bibfnamefont {G.}~\bibnamefont
  {Cabass}}, \bibinfo {author} {\bibfnamefont {E.}~\bibnamefont {Pajer}}, \
  and\ \bibinfo {author} {\bibfnamefont {F.}~\bibnamefont {Schmidt}},\ }\href
  {\doibase 10.1088/1475-7516/2018/09/003} {\bibfield  {journal} {\bibinfo
  {journal} {JCAP}\ }\textbf {\bibinfo {volume} {09}},\ \bibinfo {pages} {003}
  (\bibinfo {year} {2018})},\ \Eprint {http://arxiv.org/abs/1804.07295}
  {arXiv:1804.07295 [astro-ph.CO]} \BibitemShut {NoStop}%
\bibitem [{\citenamefont {Duaso~Pueyo}\ and\ \citenamefont
  {Pajer}(2023)}]{DuasoPueyo:2023viy}%
  \BibitemOpen
  \bibfield  {author} {\bibinfo {author} {\bibfnamefont {C.}~\bibnamefont
  {Duaso~Pueyo}}\ and\ \bibinfo {author} {\bibfnamefont {E.}~\bibnamefont
  {Pajer}},\ }\href@noop {} {\  (\bibinfo {year} {2023})},\ \Eprint
  {http://arxiv.org/abs/2311.01395} {arXiv:2311.01395 [hep-th]} \BibitemShut
  {NoStop}%
\bibitem [{\citenamefont {Creminelli}\ \emph {et~al.}(2024)\citenamefont
  {Creminelli}, \citenamefont {Renaux-Petel}, \citenamefont {Tambalo},\ and\
  \citenamefont {Yingcharoenrat}}]{Creminelli:2024cge}%
  \BibitemOpen
  \bibfield  {author} {\bibinfo {author} {\bibfnamefont {P.}~\bibnamefont
  {Creminelli}}, \bibinfo {author} {\bibfnamefont {S.}~\bibnamefont
  {Renaux-Petel}}, \bibinfo {author} {\bibfnamefont {G.}~\bibnamefont
  {Tambalo}}, \ and\ \bibinfo {author} {\bibfnamefont {V.}~\bibnamefont
  {Yingcharoenrat}},\ }\href {\doibase 10.1007/JHEP03(2024)010} {\bibfield
  {journal} {\bibinfo  {journal} {JHEP}\ }\textbf {\bibinfo {volume} {03}},\
  \bibinfo {pages} {010} (\bibinfo {year} {2024})},\ \Eprint
  {http://arxiv.org/abs/2401.10212} {arXiv:2401.10212 [hep-th]} \BibitemShut
  {NoStop}%
\bibitem [{\citenamefont {Dong}\ \emph {et~al.}(2011)\citenamefont {Dong},
  \citenamefont {Horn}, \citenamefont {Silverstein},\ and\ \citenamefont
  {Westphal}}]{Dong:2010in}%
  \BibitemOpen
  \bibfield  {author} {\bibinfo {author} {\bibfnamefont {X.}~\bibnamefont
  {Dong}}, \bibinfo {author} {\bibfnamefont {B.}~\bibnamefont {Horn}}, \bibinfo
  {author} {\bibfnamefont {E.}~\bibnamefont {Silverstein}}, \ and\ \bibinfo
  {author} {\bibfnamefont {A.}~\bibnamefont {Westphal}},\ }\href {\doibase
  10.1103/PhysRevD.84.026011} {\bibfield  {journal} {\bibinfo  {journal} {Phys.
  Rev. D}\ }\textbf {\bibinfo {volume} {84}},\ \bibinfo {pages} {026011}
  (\bibinfo {year} {2011})},\ \Eprint {http://arxiv.org/abs/1011.4521}
  {arXiv:1011.4521 [hep-th]} \BibitemShut {NoStop}%
\bibitem [{\citenamefont {Flauger}\ \emph {et~al.}(2017)\citenamefont
  {Flauger}, \citenamefont {Mirbabayi}, \citenamefont {Senatore},\ and\
  \citenamefont {Silverstein}}]{Flauger:2016idt}%
  \BibitemOpen
  \bibfield  {author} {\bibinfo {author} {\bibfnamefont {R.}~\bibnamefont
  {Flauger}}, \bibinfo {author} {\bibfnamefont {M.}~\bibnamefont {Mirbabayi}},
  \bibinfo {author} {\bibfnamefont {L.}~\bibnamefont {Senatore}}, \ and\
  \bibinfo {author} {\bibfnamefont {E.}~\bibnamefont {Silverstein}},\ }\href
  {\doibase 10.1088/1475-7516/2017/10/058} {\bibfield  {journal} {\bibinfo
  {journal} {JCAP}\ }\textbf {\bibinfo {volume} {10}},\ \bibinfo {pages} {058}
  (\bibinfo {year} {2017})},\ \Eprint {http://arxiv.org/abs/1606.00513}
  {arXiv:1606.00513 [hep-th]} \BibitemShut {NoStop}%
\bibitem [{\citenamefont {Pedro}\ and\ \citenamefont
  {Westphal}(2019)}]{Pedro:2019klo}%
  \BibitemOpen
  \bibfield  {author} {\bibinfo {author} {\bibfnamefont {F.~G.}\ \bibnamefont
  {Pedro}}\ and\ \bibinfo {author} {\bibfnamefont {A.}~\bibnamefont
  {Westphal}},\ }\href@noop {} {\  (\bibinfo {year} {2019})},\ \Eprint
  {http://arxiv.org/abs/1909.08100} {arXiv:1909.08100 [hep-th]} \BibitemShut
  {NoStop}%
\bibitem [{\citenamefont {Bhattacharya}\ and\ \citenamefont
  {Zavala}(2023)}]{Bhattacharya:2022fze}%
  \BibitemOpen
  \bibfield  {author} {\bibinfo {author} {\bibfnamefont {S.}~\bibnamefont
  {Bhattacharya}}\ and\ \bibinfo {author} {\bibfnamefont {I.}~\bibnamefont
  {Zavala}},\ }\href {\doibase 10.1088/1475-7516/2023/04/065} {\bibfield
  {journal} {\bibinfo  {journal} {JCAP}\ }\textbf {\bibinfo {volume} {04}},\
  \bibinfo {pages} {065} (\bibinfo {year} {2023})},\ \Eprint
  {http://arxiv.org/abs/2205.06065} {arXiv:2205.06065 [astro-ph.CO]}
  \BibitemShut {NoStop}%
\bibitem [{\citenamefont {Chakraborty}\ \emph {et~al.}(2019)\citenamefont
  {Chakraborty}, \citenamefont {Chiovoloni}, \citenamefont {Loaiza-Brito},
  \citenamefont {Niz},\ and\ \citenamefont {Zavala}}]{Chakraborty:2019dfh}%
  \BibitemOpen
  \bibfield  {author} {\bibinfo {author} {\bibfnamefont {D.}~\bibnamefont
  {Chakraborty}}, \bibinfo {author} {\bibfnamefont {R.}~\bibnamefont
  {Chiovoloni}}, \bibinfo {author} {\bibfnamefont {O.}~\bibnamefont
  {Loaiza-Brito}}, \bibinfo {author} {\bibfnamefont {G.}~\bibnamefont {Niz}}, \
  and\ \bibinfo {author} {\bibfnamefont {I.}~\bibnamefont {Zavala}},\
  }\href@noop {} {\  (\bibinfo {year} {2019})},\ \Eprint
  {http://arxiv.org/abs/1908.09797} {arXiv:1908.09797 [hep-th]} \BibitemShut
  {NoStop}%
\bibitem [{\citenamefont {Cespedes}\ \emph {et~al.}(2012)\citenamefont
  {Cespedes}, \citenamefont {Atal},\ and\ \citenamefont
  {Palma}}]{Cespedes:2012hu}%
  \BibitemOpen
  \bibfield  {author} {\bibinfo {author} {\bibfnamefont {S.}~\bibnamefont
  {Cespedes}}, \bibinfo {author} {\bibfnamefont {V.}~\bibnamefont {Atal}}, \
  and\ \bibinfo {author} {\bibfnamefont {G.~A.}\ \bibnamefont {Palma}},\ }\href
  {\doibase 10.1088/1475-7516/2012/05/008} {\bibfield  {journal} {\bibinfo
  {journal} {JCAP}\ }\textbf {\bibinfo {volume} {05}},\ \bibinfo {pages} {008}
  (\bibinfo {year} {2012})},\ \Eprint {http://arxiv.org/abs/1201.4848}
  {arXiv:1201.4848 [hep-th]} \BibitemShut {NoStop}%
\bibitem [{\citenamefont {Achucarro}\ \emph
  {et~al.}(2012{\natexlab{b}})\citenamefont {Achucarro}, \citenamefont {Atal},
  \citenamefont {Cespedes}, \citenamefont {Gong}, \citenamefont {Palma},\ and\
  \citenamefont {Patil}}]{Achucarro:2012yr}%
  \BibitemOpen
  \bibfield  {author} {\bibinfo {author} {\bibfnamefont {A.}~\bibnamefont
  {Achucarro}}, \bibinfo {author} {\bibfnamefont {V.}~\bibnamefont {Atal}},
  \bibinfo {author} {\bibfnamefont {S.}~\bibnamefont {Cespedes}}, \bibinfo
  {author} {\bibfnamefont {J.-O.}\ \bibnamefont {Gong}}, \bibinfo {author}
  {\bibfnamefont {G.~A.}\ \bibnamefont {Palma}}, \ and\ \bibinfo {author}
  {\bibfnamefont {S.~P.}\ \bibnamefont {Patil}},\ }\href {\doibase
  10.1103/PhysRevD.86.121301} {\bibfield  {journal} {\bibinfo  {journal} {Phys.
  Rev. D}\ }\textbf {\bibinfo {volume} {86}},\ \bibinfo {pages} {121301}
  (\bibinfo {year} {2012}{\natexlab{b}})},\ \Eprint
  {http://arxiv.org/abs/1205.0710} {arXiv:1205.0710 [hep-th]} \BibitemShut
  {NoStop}%
\bibitem [{\citenamefont {Pinol}\ \emph {et~al.}(2023)\citenamefont {Pinol},
  \citenamefont {Renaux-Petel},\ and\ \citenamefont {Werth}}]{Pinol:2023oux}%
  \BibitemOpen
  \bibfield  {author} {\bibinfo {author} {\bibfnamefont {L.}~\bibnamefont
  {Pinol}}, \bibinfo {author} {\bibfnamefont {S.}~\bibnamefont {Renaux-Petel}},
  \ and\ \bibinfo {author} {\bibfnamefont {D.}~\bibnamefont {Werth}},\
  }\href@noop {} {\  (\bibinfo {year} {2023})},\ \Eprint
  {http://arxiv.org/abs/2312.06559} {arXiv:2312.06559 [astro-ph.CO]}
  \BibitemShut {NoStop}%
\bibitem [{\citenamefont {Werth}\ \emph {et~al.}(2024)\citenamefont {Werth},
  \citenamefont {Pinol},\ and\ \citenamefont {Renaux-Petel}}]{Werth:2023pfl}%
  \BibitemOpen
  \bibfield  {author} {\bibinfo {author} {\bibfnamefont {D.}~\bibnamefont
  {Werth}}, \bibinfo {author} {\bibfnamefont {L.}~\bibnamefont {Pinol}}, \ and\
  \bibinfo {author} {\bibfnamefont {S.}~\bibnamefont {Renaux-Petel}},\ }\href
  {\doibase 10.1103/PhysRevLett.133.141002} {\bibfield  {journal} {\bibinfo
  {journal} {Phys. Rev. Lett.}\ }\textbf {\bibinfo {volume} {133}},\ \bibinfo
  {pages} {141002} (\bibinfo {year} {2024})},\ \Eprint
  {http://arxiv.org/abs/2302.00655} {arXiv:2302.00655 [hep-th]} \BibitemShut
  {NoStop}%
\bibitem [{\citenamefont {Balasubramanian}\ \emph {et~al.}(2005)\citenamefont
  {Balasubramanian}, \citenamefont {Berglund}, \citenamefont {Conlon},\ and\
  \citenamefont {Quevedo}}]{Balasubramanian:2005zx}%
  \BibitemOpen
  \bibfield  {author} {\bibinfo {author} {\bibfnamefont {V.}~\bibnamefont
  {Balasubramanian}}, \bibinfo {author} {\bibfnamefont {P.}~\bibnamefont
  {Berglund}}, \bibinfo {author} {\bibfnamefont {J.~P.}\ \bibnamefont
  {Conlon}}, \ and\ \bibinfo {author} {\bibfnamefont {F.}~\bibnamefont
  {Quevedo}},\ }\href {\doibase 10.1088/1126-6708/2005/03/007} {\bibfield
  {journal} {\bibinfo  {journal} {JHEP}\ }\textbf {\bibinfo {volume} {03}},\
  \bibinfo {pages} {007} (\bibinfo {year} {2005})},\ \Eprint
  {http://arxiv.org/abs/hep-th/0502058} {arXiv:hep-th/0502058} \BibitemShut
  {NoStop}%
\bibitem [{\citenamefont {Wang}\ and\ \citenamefont
  {Zhang}(2025)}]{ResonantCollider}%
  \BibitemOpen
  \bibfield  {author} {\bibinfo {author} {\bibfnamefont {D.-G.}\ \bibnamefont
  {Wang}}\ and\ \bibinfo {author} {\bibfnamefont {B.}~\bibnamefont {Zhang}},\
  }\href {\doibase 10.1007/JHEP09(2025)122} {\bibfield  {journal} {\bibinfo
  {journal} {JHEP}\ }\textbf {\bibinfo {volume} {09}},\ \bibinfo {pages} {122}
  (\bibinfo {year} {2025})},\ \Eprint {http://arxiv.org/abs/2505.19066}
  {arXiv:2505.19066 [hep-th]} \BibitemShut {NoStop}%
\bibitem [{\citenamefont {Chen}(2012)}]{Chen:2011zf}%
  \BibitemOpen
  \bibfield  {author} {\bibinfo {author} {\bibfnamefont {X.}~\bibnamefont
  {Chen}},\ }\href {\doibase 10.1088/1475-7516/2012/01/038} {\bibfield
  {journal} {\bibinfo  {journal} {JCAP}\ }\textbf {\bibinfo {volume} {01}},\
  \bibinfo {pages} {038} (\bibinfo {year} {2012})},\ \Eprint
  {http://arxiv.org/abs/1104.1323} {arXiv:1104.1323 [hep-th]} \BibitemShut
  {NoStop}%
\bibitem [{\citenamefont {Shiu}\ and\ \citenamefont {Xu}(2011)}]{Shiu:2011qw}%
  \BibitemOpen
  \bibfield  {author} {\bibinfo {author} {\bibfnamefont {G.}~\bibnamefont
  {Shiu}}\ and\ \bibinfo {author} {\bibfnamefont {J.}~\bibnamefont {Xu}},\
  }\href {\doibase 10.1103/PhysRevD.84.103509} {\bibfield  {journal} {\bibinfo
  {journal} {Phys. Rev. D}\ }\textbf {\bibinfo {volume} {84}},\ \bibinfo
  {pages} {103509} (\bibinfo {year} {2011})},\ \Eprint
  {http://arxiv.org/abs/1108.0981} {arXiv:1108.0981 [hep-th]} \BibitemShut
  {NoStop}%
\bibitem [{\citenamefont {Gao}\ \emph {et~al.}(2012)\citenamefont {Gao},
  \citenamefont {Langlois},\ and\ \citenamefont {Mizuno}}]{Gao:2012uq}%
  \BibitemOpen
  \bibfield  {author} {\bibinfo {author} {\bibfnamefont {X.}~\bibnamefont
  {Gao}}, \bibinfo {author} {\bibfnamefont {D.}~\bibnamefont {Langlois}}, \
  and\ \bibinfo {author} {\bibfnamefont {S.}~\bibnamefont {Mizuno}},\ }\href
  {\doibase 10.1088/1475-7516/2012/10/040} {\bibfield  {journal} {\bibinfo
  {journal} {JCAP}\ }\textbf {\bibinfo {volume} {10}},\ \bibinfo {pages} {040}
  (\bibinfo {year} {2012})},\ \Eprint {http://arxiv.org/abs/1205.5275}
  {arXiv:1205.5275 [hep-th]} \BibitemShut {NoStop}%
\bibitem [{\citenamefont {Chen}\ \emph {et~al.}(2015)\citenamefont {Chen},
  \citenamefont {Namjoo},\ and\ \citenamefont {Wang}}]{Chen:2014cwa}%
  \BibitemOpen
  \bibfield  {author} {\bibinfo {author} {\bibfnamefont {X.}~\bibnamefont
  {Chen}}, \bibinfo {author} {\bibfnamefont {M.~H.}\ \bibnamefont {Namjoo}}, \
  and\ \bibinfo {author} {\bibfnamefont {Y.}~\bibnamefont {Wang}},\ }\href
  {\doibase 10.1088/1475-7516/2015/02/027} {\bibfield  {journal} {\bibinfo
  {journal} {JCAP}\ }\textbf {\bibinfo {volume} {02}},\ \bibinfo {pages} {027}
  (\bibinfo {year} {2015})},\ \Eprint {http://arxiv.org/abs/1411.2349}
  {arXiv:1411.2349 [astro-ph.CO]} \BibitemShut {NoStop}%
\bibitem [{\citenamefont {Gong}\ and\ \citenamefont
  {Tanaka}(2011)}]{Gong:2011uw}%
  \BibitemOpen
  \bibfield  {author} {\bibinfo {author} {\bibfnamefont {J.-O.}\ \bibnamefont
  {Gong}}\ and\ \bibinfo {author} {\bibfnamefont {T.}~\bibnamefont {Tanaka}},\
  }\href {\doibase 10.1088/1475-7516/2012/02/E01,
  10.1088/1475-7516/2011/03/015} {\bibfield  {journal} {\bibinfo  {journal}
  {JCAP}\ }\textbf {\bibinfo {volume} {1103}},\ \bibinfo {pages} {015}
  (\bibinfo {year} {2011})},\ \bibinfo {note} {[Erratum: JCAP1202,E01(2012)]},\
  \Eprint {http://arxiv.org/abs/1101.4809} {arXiv:1101.4809 [astro-ph.CO]}
  \BibitemShut {NoStop}%
\bibitem [{\citenamefont {Jazayeri}\ \emph {et~al.}(2025)\citenamefont
  {Jazayeri}, \citenamefont {Tong},\ and\ \citenamefont
  {Zhu}}]{Jazayeri:2025vlv}%
  \BibitemOpen
  \bibfield  {author} {\bibinfo {author} {\bibfnamefont {S.}~\bibnamefont
  {Jazayeri}}, \bibinfo {author} {\bibfnamefont {X.}~\bibnamefont {Tong}}, \
  and\ \bibinfo {author} {\bibfnamefont {Y.}~\bibnamefont {Zhu}},\ }\href@noop
  {} {\  (\bibinfo {year} {2025})},\ \Eprint {http://arxiv.org/abs/2511.00152}
  {arXiv:2511.00152 [hep-th]} \BibitemShut {NoStop}%
\bibitem [{\citenamefont {Cheung}\ \emph {et~al.}(2008)\citenamefont {Cheung},
  \citenamefont {Creminelli}, \citenamefont {Fitzpatrick}, \citenamefont
  {Kaplan},\ and\ \citenamefont {Senatore}}]{Cheung:2007st}%
  \BibitemOpen
  \bibfield  {author} {\bibinfo {author} {\bibfnamefont {C.}~\bibnamefont
  {Cheung}}, \bibinfo {author} {\bibfnamefont {P.}~\bibnamefont {Creminelli}},
  \bibinfo {author} {\bibfnamefont {A.~L.}\ \bibnamefont {Fitzpatrick}},
  \bibinfo {author} {\bibfnamefont {J.}~\bibnamefont {Kaplan}}, \ and\ \bibinfo
  {author} {\bibfnamefont {L.}~\bibnamefont {Senatore}},\ }\href {\doibase
  10.1088/1126-6708/2008/03/014} {\bibfield  {journal} {\bibinfo  {journal}
  {JHEP}\ }\textbf {\bibinfo {volume} {03}},\ \bibinfo {pages} {014} (\bibinfo
  {year} {2008})},\ \Eprint {http://arxiv.org/abs/0709.0293} {arXiv:0709.0293
  [hep-th]} \BibitemShut {NoStop}%
\bibitem [{\citenamefont {Garcia-Saenz}\ \emph {et~al.}(2019)\citenamefont
  {Garcia-Saenz}, \citenamefont {Pinol},\ and\ \citenamefont
  {Renaux-Petel}}]{Garcia-Saenz:2019njm}%
  \BibitemOpen
  \bibfield  {author} {\bibinfo {author} {\bibfnamefont {S.}~\bibnamefont
  {Garcia-Saenz}}, \bibinfo {author} {\bibfnamefont {L.}~\bibnamefont {Pinol}},
  \ and\ \bibinfo {author} {\bibfnamefont {S.}~\bibnamefont {Renaux-Petel}},\
  }\href@noop {} {\  (\bibinfo {year} {2019})},\ \Eprint
  {http://arxiv.org/abs/1907.10403} {arXiv:1907.10403 [hep-th]} \BibitemShut
  {NoStop}%
\bibitem [{\citenamefont {Qin}\ and\ \citenamefont
  {Xianyu}(2023)}]{Qin:2023ejc}%
  \BibitemOpen
  \bibfield  {author} {\bibinfo {author} {\bibfnamefont {Z.}~\bibnamefont
  {Qin}}\ and\ \bibinfo {author} {\bibfnamefont {Z.-Z.}\ \bibnamefont
  {Xianyu}},\ }\href {\doibase 10.1007/JHEP07(2023)001} {\bibfield  {journal}
  {\bibinfo  {journal} {JHEP}\ }\textbf {\bibinfo {volume} {07}},\ \bibinfo
  {pages} {001} (\bibinfo {year} {2023})},\ \Eprint
  {http://arxiv.org/abs/2301.07047} {arXiv:2301.07047 [hep-th]} \BibitemShut
  {NoStop}%
\bibitem [{\citenamefont {Akrami}\ \emph {et~al.}(2020)\citenamefont {Akrami}
  \emph {et~al.}}]{Planck:2018jri}%
  \BibitemOpen
  \bibfield  {author} {\bibinfo {author} {\bibfnamefont {Y.}~\bibnamefont
  {Akrami}} \emph {et~al.} (\bibinfo {collaboration} {Planck}),\ }\href
  {\doibase 10.1051/0004-6361/201833887} {\bibfield  {journal} {\bibinfo
  {journal} {Astron. Astrophys.}\ }\textbf {\bibinfo {volume} {641}},\ \bibinfo
  {pages} {A10} (\bibinfo {year} {2020})},\ \Eprint
  {http://arxiv.org/abs/1807.06211} {arXiv:1807.06211 [astro-ph.CO]}
  \BibitemShut {NoStop}%
\bibitem [{\citenamefont {Sohn}\ \emph {et~al.}(2023)\citenamefont {Sohn},
  \citenamefont {Fergusson},\ and\ \citenamefont {Shellard}}]{Sohn:2023fte}%
  \BibitemOpen
  \bibfield  {author} {\bibinfo {author} {\bibfnamefont {W.}~\bibnamefont
  {Sohn}}, \bibinfo {author} {\bibfnamefont {J.~R.}\ \bibnamefont {Fergusson}},
  \ and\ \bibinfo {author} {\bibfnamefont {E.~P.~S.}\ \bibnamefont
  {Shellard}},\ }\href {\doibase 10.1103/PhysRevD.108.063504} {\bibfield
  {journal} {\bibinfo  {journal} {Phys. Rev. D}\ }\textbf {\bibinfo {volume}
  {108}},\ \bibinfo {pages} {063504} (\bibinfo {year} {2023})},\ \Eprint
  {http://arxiv.org/abs/2305.14646} {arXiv:2305.14646 [astro-ph.CO]}
  \BibitemShut {NoStop}%
\bibitem [{\citenamefont {Sohn}\ \emph {et~al.}(2024)\citenamefont {Sohn},
  \citenamefont {Wang}, \citenamefont {Fergusson},\ and\ \citenamefont
  {Shellard}}]{Sohn:2024xzd}%
  \BibitemOpen
  \bibfield  {author} {\bibinfo {author} {\bibfnamefont {W.}~\bibnamefont
  {Sohn}}, \bibinfo {author} {\bibfnamefont {D.-G.}\ \bibnamefont {Wang}},
  \bibinfo {author} {\bibfnamefont {J.~R.}\ \bibnamefont {Fergusson}}, \ and\
  \bibinfo {author} {\bibfnamefont {E.~P.~S.}\ \bibnamefont {Shellard}},\
  }\href {\doibase 10.1088/1475-7516/2024/09/016} {\bibfield  {journal}
  {\bibinfo  {journal} {JCAP}\ }\textbf {\bibinfo {volume} {09}},\ \bibinfo
  {pages} {016} (\bibinfo {year} {2024})},\ \Eprint
  {http://arxiv.org/abs/2404.07203} {arXiv:2404.07203 [astro-ph.CO]}
  \BibitemShut {NoStop}%
\bibitem [{\citenamefont {Suman}\ \emph
  {et~al.}(2025{\natexlab{a}})\citenamefont {Suman}, \citenamefont {Wang},
  \citenamefont {Sohn}, \citenamefont {Fergusson},\ and\ \citenamefont
  {Shellard}}]{Suman:2025vuf}%
  \BibitemOpen
  \bibfield  {author} {\bibinfo {author} {\bibfnamefont {P.}~\bibnamefont
  {Suman}}, \bibinfo {author} {\bibfnamefont {D.-G.}\ \bibnamefont {Wang}},
  \bibinfo {author} {\bibfnamefont {W.}~\bibnamefont {Sohn}}, \bibinfo {author}
  {\bibfnamefont {J.~R.}\ \bibnamefont {Fergusson}}, \ and\ \bibinfo {author}
  {\bibfnamefont {E.~P.~S.}\ \bibnamefont {Shellard}},\ }\href@noop {} {\
  (\bibinfo {year} {2025}{\natexlab{a}})},\ \Eprint
  {http://arxiv.org/abs/2511.17500} {arXiv:2511.17500 [astro-ph.CO]}
  \BibitemShut {NoStop}%
\bibitem [{\citenamefont {Suman}\ \emph
  {et~al.}(2025{\natexlab{b}})\citenamefont {Suman}, \citenamefont {Wang},
  \citenamefont {Sohn}, \citenamefont {Fergusson},\ and\ \citenamefont
  {Shellard}}]{Suman:2025tpv}%
  \BibitemOpen
  \bibfield  {author} {\bibinfo {author} {\bibfnamefont {P.}~\bibnamefont
  {Suman}}, \bibinfo {author} {\bibfnamefont {D.-G.}\ \bibnamefont {Wang}},
  \bibinfo {author} {\bibfnamefont {W.}~\bibnamefont {Sohn}}, \bibinfo {author}
  {\bibfnamefont {J.~R.}\ \bibnamefont {Fergusson}}, \ and\ \bibinfo {author}
  {\bibfnamefont {E.~P.~S.}\ \bibnamefont {Shellard}},\ }\href@noop {} {\
  (\bibinfo {year} {2025}{\natexlab{b}})},\ \Eprint
  {http://arxiv.org/abs/2512.22085} {arXiv:2512.22085 [astro-ph.CO]}
  \BibitemShut {NoStop}%
\end{thebibliography}%
\end{document}